\documentclass[twocolumn,twocolappendix,trackchanges]{aastex631}

\usepackage{amsmath}
\usepackage{enumerate}
\usepackage{threeparttable}
\usepackage{lipsum}
\received{}
\revised{}
\accepted{}
\submitjournal{ApJ}

\newcommand{\Msun}{{\rm  M_{\odot}}}

\newcommand{\kms}{\rm km\;s^{-1}}

\newcommand{\micro}{\rm \mu m}
\newcommand{\obs}{\rm {\theta_{obs}}}
\graphicspath{{./}{figures/}}

\begin{document}

\title{Probing dynamics and thermal properties inside molecular tori with CO rovibrational absorption lines}
\author[0000-0002-5012-6707]{Kosei Matsumoto}
\email{koseim@ir.isas.jaxa.jp}
\affiliation{{\rm Department of Physics, Graduate School of Science, The University of Tokyo, 7-3-1 Hongo, Bunkyo-ku, Tokyo 113-0033, Japan}}
\affiliation{{\rm Institute of Space and Astronautical Science, Japan Aerospace Exploration Agency, 3-1-1 Yoshinodai, Chuo-ku, Sagamihara, Kanagawa 252-5210, Japan}}
\affiliation{\rm{Sterrenkundig Observatorium, Universiteit Gent, Krijgslaan 281 S9, B-9000 Gent, Belgium}}

\author[0000-0002-6660-9375]{Takao Nakagawa}
\affiliation{{\rm Institute of Space and Astronautical Science, Japan Aerospace Exploration Agency, 3-1-1 Yoshinodai, Chuo-ku, Sagamihara,
Kanagawa 252-5210, Japan}}

\author[0000-0002-8779-8486]{Keiichi Wada}
\affiliation{{\rm Kagoshima University, Graduate School of Science and Engineering, Kagoshima 890-0065, Japan}}
\affiliation{{\rm Ehime University, Research Center for Space and Cosmic Evolution, Matsuyama 790-8577, Japan}}
\affiliation{{\rm Hokkaido University, Faculty of Science, Sapporo 060-0810, Japan}}


\author[0000-0002-9850-6290]{Shunsuke Baba}
\affiliation{{\rm Kagoshima University, Graduate School of Science and Engineering, Kagoshima 890-0065, Japan}}


\author[0000-0002-1765-7012]{Shusuke Onishi}
\affiliation{{\rm Department of Physics, Graduate School of Science, The University of Tokyo, 7-3-1 Hongo, Bunkyo-ku, Tokyo 113-0033, Japan}}
\affiliation{{\rm Institute of Space and Astronautical Science, Japan Aerospace Exploration Agency, 3-1-1 Yoshinodai, Chuo-ku, Sagamihara, Kanagawa 252-5210, Japan}}

\author{Taisei Uzuo}
\affiliation{{\rm Kagoshima University, Graduate School of Science and Engineering, Kagoshima 890-0065, Japan}}

\author{Naoki Isobe}
\affiliation{{\rm Institute of Space and Astronautical Science, Japan Aerospace Exploration Agency, 3-1-1 Yoshinodai, Chuo-ku, Sagamihara,
Kanagawa 252-5210, Japan}}

\author[0000-0003-0548-1766]{Yuki Kudoh}
\affiliation{{\rm Kagoshima University, Graduate School of Science and Engineering, Kagoshima 890-0065, Japan}}
\affiliation{{\rm National Astronomical Observatory of Japan, Mitaka 181-8588, Japan}}


\begin{abstract}
A recent hydrodynamic model, ``radiation-driven fountain model" \citep{Wada2016}, presented a dynamical picture that active galactic nuclei (AGNs) tori sustain their geometrical thickness by gas circulation around AGNs, and previous papers confirmed that this picture is consistent with multi-wavelength observations of nearby Seyfert galaxies.
Recent near-infrared observations implied that CO rovibrational absorption lines ($\Delta J=\pm1$, $v=0-1$, $\lambda \sim 4.7\, \micron$) could probe physical properties of the inside tori.
However, the origin of the CO absorption lines has been under debate.
In this paper, we investigate the origin of the absorption lines and conditions for detecting them by performing line radiative transfer calculations based on the radiation-driven fountain model. 
We find that CO rovibrational absorption lines are detected at inclination angles $\obs = 50-80 \, ^{\circ}$.
At the inclination angle $\obs = 77 \, ^{\circ}$, we observe multi-velocity components: inflow ($v_\mathrm{LOS}=30 \,\kms$), systemic  ($v_\mathrm{LOS}=0 \,\kms$), and outflows ($v_\mathrm{LOS}=-75,\, -95,$ and $-105 \,\kms$).
The inflow and outflow components ($v_\mathrm{LOS}= 30$ and $-95 \, \kms$) are collisionally excited at the excitation temperature of $186$ and $380$ K up to $J=12$ and $4$, respectively.
The inflow and outflow components originate from the accreting gas on the equatorial plane at $1.5$ pc from the AGN center and the outflowing gas driven by AGN radiation pressure at $1.0$ pc, respectively.
These results suggest that CO rovibrational absorption lines can provide us with the velocities and kinetic temperatures of the inflow and outflow in the inner a-few-pc regions of AGN tori, and the observations can probe the gas circulation inside the tori.


\end{abstract}



\section{Introduction} \label{sec:intro}
Active galactic nuclei (AGNs) are classified into two types: objects with broad emission lines in optical wavelength (i.e., type 1 AGNs) and objects without broad emission lines (i.e., type 2 AGNs).
Following spectropolarimetric observations, \citet{Miller1983,Antonucci1985,Antonucci1993}, proposed an AGN unified model, in which the AGN types are intrinsically the same.
Objects near the face-on geometry showed the broad emission lines emitted from the vicinity of the AGNs, but those at the edge-on geometry did not show broad emission lines because they are obscured by geometrically thick structures, molecular tori. 
Thus, the inclination angle to the torus is an essential factor to explain the difference between type 1 and type 2 AGNs.
However, the formation mechanism of the geometrical thickness of AGN tori is under debate.
Hence, it is important to understand the internal structures of AGN tori.

In the last decade, the Atacama Large Millimeter/submillimeter Array (ALMA) has discovered obscuring structures, like molecular tori, in nearby AGNs \citep{Garcia-Burillo2016,Combes2019A&A...623A..79C,Imanishi2018a,Imanishi2020}.
However, even ALMA has not been able to spatially resolve the internal structures of molecular tori.
Therefore, it is important to predict synthetic observational properties, such as spectral shapes, based on realistic theoretical models, and then compare the synthetic properties with the actual observational properties.
These predictions are essential to interpret observational results because the distribution of the physical values along the line of sight and the radiation processes are not obtained from observations.


\citet{Wada2012} developed a hydrodynamic model, the ``radiation-driven fountain model" with a radiative hydrodynamic simulation \citep[see also][]{Wada2009ApJ...702...63W,Namekata2016,Chan2017ApJ...843...58C,Dorodnitsyn2017ApJ...842...43D,Williamson2020}, and he suggested that inflowing gas into the center of AGN was driven by the radiation from the accretion disk and converted to outflow gas.
Some of the outflow gas returns to the disk plane and eventually accretes toward the central black hole.
This circulation of the gas naturally forms the geometrically thick torus around the AGN.
It is also confirmed that the quasi-steady torus explains the differences in the spectral energy distributions (SEDs) between type 1 and type 2 AGNs depending on the viewing angle \citep{Schartmann2014}.

\citet{Wada2016} considered the X-ray dominated region (XDR) chemistry with  the radiation-driven fountain model for the nearby type 2 AGN, the Circinus galaxy, and they found that the circumnuclear region was filled with multi-phase gases: ionized gas, atomic gas, and molecular gas.
Based on this model, synthetic observations have been performed and compared with actual observations.
\citet{Wada2018a} predicted CO rotational emission lines and [CI] emission lines, and \citet{Izumi2018} confirmed that the turbulent structure of the entire torus in the observation was consistent with the theoretical prediction.
Besides, \citet{Wada2018b} studied emission lines from the ionized gas based on the same model, and they explained the properties of a narrow-line region.

Suppose the radiation-driven fountain model explains some multi-wavelength properties of the tori, then the remaining issue will be how we can know the internal structures of the tori.
We can use molecular absorption lines to know the physical condition of the molecular gas on the line of sight.
\citet{Uzuo2021} investigated the CO rotational absorption lines in submillimeter wavelength based on the model.
The detected CO absorption line profiles for the rotational levels $J=7–6$ or higher were affected by those of the CO emission lines, especially if the beam size was not fine enough, and it was difficult to derive the kinematics and level population of the molecular gas inside the torus.
Another possible way to explore the inside of the torus is by observation with CO rovibrational absorption lines (R ($J$): $J$ ($v=0$) $\rightarrow$ $J+1$ ($v=1$), P($J$): $J$ ($v=0$) $\rightarrow$ $J-1$ ($v=1$), wavelength $\lambda \sim 4.7\, \micro$) in the near-infrared wavelength \citep[see][]{Spoon2004,Shirahata2013}.
In spectroscopy around the wavelength $4.7\, \micro$, many absorption lines arising from different rotational levels are observed simultaneously.
Since the lines cover several rotational levels, we can determine the level population of CO molecules and estimate the physical conditions on the line of sight.

\citet{Baba2018} investigated CO absorption lines in $10$ galaxies with low-dispersion spectroscopy of AKARI and Spitzer, and they estimated that the excitation temperatures of the gas were high at $200-500$ K.
Since the derived gas temperature is much higher than the typical gas temperature ($T\sim 10 $ K) in star-forming regions, they suggested that these gases originated from the inner region of the AGN tori. 
\citet{Onishi2021} decomposed CO absorption lines into multiple velocity components: inflow, systemic, and outflow ($v_\mathrm{LOS}=+100,\, 0,$ and $ -160 \ \kms$, respectively) in IRAS $08572+3915$, with high-dispersion spectroscopy of the Subaru telescope \citep[see also][]{Geballe2006,Shirahata2013}.
The excitation temperatures of the inflow and outflow components were higher than $500$ K, whereas that of the systemic component was a few ten K \citep{Onishi2021}.
They inferred that these velocity components originated from different regions inside the torus. 

However, these studies depend on the assumption that the background continuum could be the thermal dust emission from the dust sublimation layer (dust temperature of $T_\mathrm{dust} = 1200-1500$ K), though the background continuum was not spatially resolved.
Thus, the following is still under debate: against what background source CO rovibrational absorption lines are observed and whether the gases traced by the CO rovibrational absorption lines originate really in the tori.
Additionally, observations alone cannot reveal the physical conditions of gas, such as density, kinetic temperature, and radiation field along the line of sight, and hence, there is always a debate on what determines the excited states of the observed CO molecules \citep{Shirahata2013,Onishi2021}.


In this paper, to discuss the origins of observed CO rovibrational absorption lines, we focus on the following four questions:
(1) what is the source of continuum radiation at the back of the CO absorption gas in near-infrared observations, especially at the wavelength $4.7\, \micron$?; (2) under what conditions are CO absorption lines detected? (e.g., conditions for CO molecules covering the background source and inclination angles at which CO absorption lines can be detected); (3) what is the origin of the gas traced by the CO rovibrational absorption lines?; (4) what determines the excited states of the observed CO molecules in molecular tori: collision or radiation?

The remainder of this paper is organized as follows.
Section \ref{sec:method} describes our input model, the radiation-driven
fountain model, and the modeling of the radiation transfer calculations for the thermal dust emission and CO lines.
Section \ref{sec:result} contains results of the dust and CO line radiative transfer calculations. 
Section \ref{sec:discuss} discusses the above four questions based on the obtained results.
Finally, Section \ref{sec:concl} gives conclusions.
\section{Methods} \label{sec:method}
\subsection{Outline}
Here, we give an outline of the flow of our method, and in subsequent sections, we describe the details of each step. 
To model CO rovibrational absorption spectra, we use the results of our hydrodynamic model, the radiation-driven fountain model \citep{Wada2016}, as inputs of postprocessing radiative transfer calculations. 
Then, we perform three-dimensional (3D) radiative transfer calculations in the following two steps.
As a first step, we perform dust radiative transfer calculations to determine the temperature distribution of different dust grains.
As the next step, we perform non-local thermodynamic equilibrium (non-LTE) line radiative transfer calculations to determine the CO level populations at each grid.
In those radiative transfer calculations, we smooth the result of hydrodynamic simulation from $256^3$ to $128^3$ grids to save computational cost.


\begin{figure*}[]
    \centering
    \includegraphics[width=2.0\columnwidth]{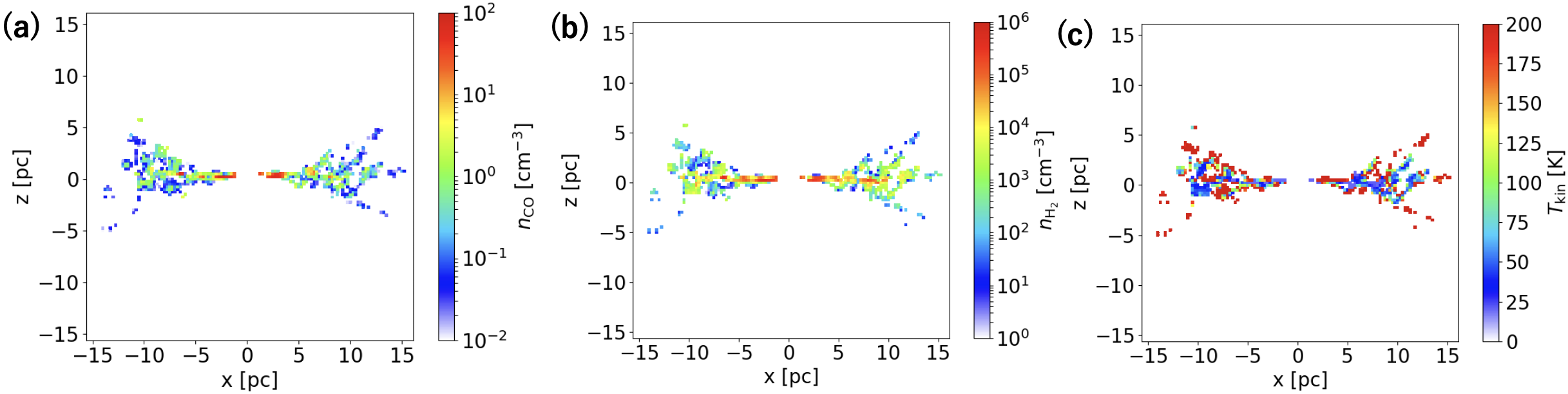}
    \caption{\label{fig:metho1}(a) and (b) show the maps of the number density of CO and $\mathrm{H_2}$ molecules, which are collision partners of CO molecules, on the x-z planes, respectively. (c) shows the map of gas kinetic temperature on the x-z plane. We note that only regions with $n_\mathrm{CO}>0.01$ cm$^{-3}$ are shown, to focus on CO molecular gas properties. } 
\end{figure*}

\subsection{Hydrodynamic model:\\ radiation-driven fountain model}\label{sec:method1}
We use a snapshot of the radiation-driven fountain model \citep{Wada2016}, which is a 3D Eulerian hydrodynamic simulation composed of uniform Cartesian $256^3 \ \mathrm{grids}$ in a box with the size of $32^3 \, \mathrm{pc^3}$. 
\citet{Wada2016} modeled a Compton-thick Seyfert AGN, known as Circinus galaxy.
The main physical parameters in this model are shown in Table \ref{tab:2-1}.
This model incorporates radiative processes of anisotropic ultraviolet (UV) radiation and isotropic X-ray radiation from the accretion disk by the ray-tracing method \citep{Netzer1987}.
The radiation pressure caused by UV radiation and heating of the X-ray radiation drive outflows around the AGN. 
Additionally, the following heating processes due to star formation activities are included: supernova heating, which is injected near the equatorial plane, and photoelectric heating in a uniform far-ultraviolet radiation field.
The radiation from the AGN and heating due to star formation activities increase the temperature of molecular gas ($T_\mathrm{kin}>100$ K) in photodissociation regions (PDR) and XDR \citep{Maloney1996}, as shown in Figure \ref{fig:metho1}.
The gas cooling function and molecular abundance are calculated using the chemical network model of \citet{Meijerink2005}, and solar metallicity is assumed.
The chemical networks include the following 25 species: $\mathrm{ H}$, $\mathrm{H_2}$, $\mathrm{  H^+}$, $\mathrm{ H^+_2}$, $\mathrm{ H^+_3}$, $\mathrm{ H^{-}}$, $\mathrm{ e^-}$, $\mathrm{ O}$, $\mathrm{ O_2}$, $\mathrm{ O^+}$, $\mathrm{ O^+_2}$, $\mathrm{ O_2H^+}$, $\mathrm{ OH}$, $\mathrm{ OH^+}$, $\mathrm{ H_2O}$, $\mathrm{ H_2O^+}$, $\mathrm{ H_3O^+}$, $\mathrm{C}$,  $\mathrm{C^+, \, CO,\, Na,\, Na^+,\, He,\, He^+, \ and\ HCO^+}$.



\begin{deluxetable}{cccc}
   \tablecaption{Main parameters in the radiation-driven fountain model by \citet{Wada2016}.}
 \label{tab:2-1}
\tablehead{
\colhead{$M_\mathrm{BH}$ ($\Msun$)}&\colhead{$M_\mathrm{gas}$ ($\Msun$)}&\colhead{$L_\mathrm{AGN}$ (erg/s) }&\colhead{$L_\mathrm{X}$ (erg/s)}
}
\decimalcolnumbers
\startdata
 $2 \times 10^6 $ & $2 \times 10^6$ &$4.9 \times 10^{43}$ & $2.8 \times 10^{42}$ 
  \enddata
\tablecomments{Column (1): Black hole mass. Column (2): Initial gas mass. Column (3): AGN luminosity. Column (4): X-ray luminosity.}
\end{deluxetable}


\subsection{3D dust radiative transfer}\label{sec:method2}
To calculate the temperature distributions of dust grains and thermal dust emissions at a wavelength of $4.7\, \micron$, we use RADMC-3D \citep{Dullemond2012}\footnote{\url{https://www.ita.uni-heidelberg.de/~dullemond/software/radmc-3d/}}, which is a 3D dust radiation transfer code that fully considers thermal dust emission, absorption, and scattering.

RADMC-3D requires modeling dust properties and external radiation source.
We use four dust species: silicate and graphite grains with sizes in $0.01\, \micro$ and $0.1\, \micro$.
Optical properties of these species (e.g., absorption opacity $\kappa_{\nu, k}$) are taken from \citet{Laor1993}.
The density of dust grains is scaled from the gas density, assuming a uniform gas-to-dust mass ratio of $0.01$. 
Then, the mass of each grain in each cell is the same as in \citet{Schartmann2005}, which assumes a typical galactic dust model for the dust size distribution.
In this dust model, the dust number density distribution of different sizes is assumed to be proportional to $a^{-3.5}$, where $a$ is the grain size (MRN  size distribution, \citet{Mathis1977}), and the dust mass in each cell is composed of $62.5\%$ silicate and $37.5 \%$ graphite.

We set the external radiation source at the center of the calculation box to represent an accretion disk, which heats dust grains, and we assume radiation equilibrium to determine the temperature of each dust grain.
The source radiates isotropically, and its SED from $10^{-3} \, \micron$ to $10^{3} \, \micron$ is taken from \citet{Schartmann2005}.



\subsection{3D non-LTE line radiative transfer}\label{sec:method3}
For the modeling of CO rovibrational absorption lines, we modify our 3D non-LTE line radiative transfer code \citep{Hogerheijde2000A&A...362..697H,Wada2005, Yamada2007ApJ...671...73Y, Wada2018a}, which calculates populations of rotational levels of molecules and makes mock spectroscopic observations in any direction of sight.
Here, the direction is specified by the inclination and azimuthal angles defined in the polar coordinate in the calculation box.
To include the effect of rovibrational transitions in our calculations, we consider both rotational and vibrational levels for rovibrational transitions.
Besides, to reproduce CO absorption lines against dust continuum self-consistently, we deal with dust absorption and thermal dust emission of independent dust grains in the line radiative transfer calculations.
Here, we do not incorporate dust scattering into the line radiative transfer calculations because the dust scattering contribution is uncritical on the radiative transfer calculations at a wavelength of $4.7\, \micron$.
We found that the scattering contribution is about $3\%$ of the total extinction using RADMC-3D.

Radiative transfer calculations for all transition lines, which include absorption and emission by N dust grains, are calculated as
\begin{linenomath}
\begin{equation}
\frac{dI_\nu}{d\tau_\nu} = -I_\nu+ S_\nu,
\label{eq:3.1}
\end{equation}
\end{linenomath}
where $I_\nu$ is the intensity; $\tau_\nu$ is the optical depth given by $d\tau_\nu \equiv (\alpha_{\nu,ul}+\sum^{N}_{k=1} \alpha_{\nu, k}) ds $, where $\alpha_{\nu,ul}$, $\alpha_{\nu, k}$, and $ds$ are the absorption coefficient of each CO level transition, that of the $k$th dust grain, and length; $S_\nu$ is the total source function, which is defined as
\begin{linenomath}
\begin{equation}
S_\nu \equiv \frac{j_{\nu,ul}+\sum^{N}_{k=1} \alpha_{\nu, k} B_{\nu}(T_k) }{\alpha_{\nu,ul} +\sum^{N}_{k=1} \alpha_{\nu, k}},
\label{eq:3.2}
\end{equation}
\end{linenomath}
where $B_{\nu}(T_k)$ and $j_{\nu,ul}$ are the intensities of the black-body radiation with the temperature $T_k$ and emissivity of each CO transition, respectively.
Here, we assume that the line profile per grid is defined as the Gaussian profile dominated by the microturbulence of $10\, \kms$ \citep{Izumi2018}.
The observed spectrum is represented as a Gaussian's superposition containing the velocity fields of the hydrodynamic simulation.

The mean intensity at each grid is obtained by solving Equation (\ref{eq:3.1}) of all incident rays into the grid and averaging the intensity of all incident rays.
These radiative transfer calculations are performed with the calculation of the statistical equilibrium equation iteratively, so that we obtain the convergent solutions.
The statistical equilibrium equation is written as
\begin{linenomath}
\begin{equation}
\begin{split}
& \Bigg[ n_{i+1}A_{i+1,i} + (n_{i+1}B_{i+1,i}-n_iB_{i,i+1})J_{i+1,i} \\
& - n_{i}A_{i,i-1}-(n_iB_{i,i-1}-n_{i-1}B_{i-1,i})J_{i,i-1}\\
& + \sum^{N_\mathrm{lev}}_{j\neq i}\big[n_jC_{ji}-n_iC_{ij}\big]\Bigg]_{v=0}\\
& + \Bigg[\sum_{i^{'}=i\pm1} \Big[n_{i^{'}}A_{{i^{'}},i} +  (n_{i^{'}}B_{{i^{'}},i}-n_iB_{i,{i^{'}}})J_{{i^{'}},i}\Big]\\
& + \sum^{N_\mathrm{lev}}_{i^{'}}\big[n_{i^{'}}C_{i^{'},i}-n_iC_{i,i^{'}}\big]\Bigg]_{v=0-1}=0,
\label{eq:equibulium_eq}
\end{split}
\end{equation}
\end{linenomath}
where $N_\mathrm{lev}$ is the number of rotational levels in the vibrational level $v=0$; $n_i$ and $n_{i^{'}}$ are the number densities of CO at the $i$th rotational level in the vibrational levels $v=0$ and $v=1$, respectively; $A$ and $B$ are the Einstein coefficients related to local radiation fields; $C$ is the collisional coefficient depending on the local gas density and kinetic temperature.
In Equation (\ref{eq:equibulium_eq}), the first term represents the transitions between rotational levels in vibrational level $v=0$, and the second term represents the rovibrational transitions between vibrational levels $v=0$ and $v=1$.
We note that we do not consider only the rotational transitions in vibrational level $v=1$, since $A$ coefficients of rovibrational transitions $v=0-1$ are much larger than those of rotational transitions in vibrational level $v=1$.
Here, we consider 42 CO energy levels ($J=0-20$, $v=0$, $1$).
We extract $A$ coefficients from the HITRAN database \citep{Rothman2013}, collisional coefficients for rotational transitions from \citet{Yang2010ApJ...718.1062Y}, and collisional coefficients for rovibrational transitions from \citet{Song2015}.
We note that the collisional transitions between rovibrational levels are not efficient because the critical density ($n_\mathrm{H_2} \sim 10^{15}$ cm$^{-3}$) is significantly high compared to the maximum gas density of our simulation ($n_\mathrm{H_2}< 1\times 10^{8}$ cm$^{-3}$).  
\section{results} \label{sec:result}
\subsection{Background source for CO rovibrational absorption lines: $4.7\, \micron$ dust continuum} \label{sec:result1}
\begin{figure}[htb!]
    \centering
    \includegraphics[width=1.0\columnwidth]{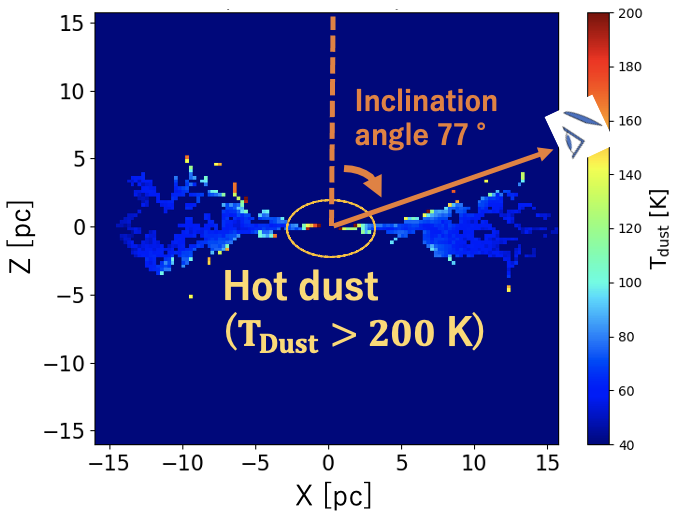}
    \caption{\label{fig:result1}Dust temperature map of the $0.1\, \micron$ graphite dust grain on the x-z plane. Hot dust ($T_\mathrm{dust}>200$ K) exists within the orange circle. We note that only the regions with $n_\mathrm{CO}>0.01$ (cm$^{-3}$) are shown. The orange arrow shows the case that the torus is observed at the inclination angle $\obs=77 \, ^{\circ}$ and azimuthal angle $\phi_\mathrm{obs}=0 \, ^{\circ}$.} 
\end{figure}
To investigate the background source of the CO rovibrational absorption lines, we explore the results of the dust radiative transfer calculations.

Figure \ref{fig:result1} shows the temperature map of the $0.1\, \micron$ graphite dust grain on the x-z plane.
Cold dust at a temperature of a few ten K exists near the equatorial plane of the torus, whereas warm dust at a temperature of $100 - 200 \, \mathrm{K}$ exists in the cone-shaped inner-wall region of the torus, where the dust is directly irradiated by the central AGN radiation.
Hot dust at a temperature higher than $200 \, \mathrm{K}$ is concentrated in the inner $1.5$-pc region of the torus.

\begin{figure}
    \centering
    \includegraphics[width=1.0\columnwidth]{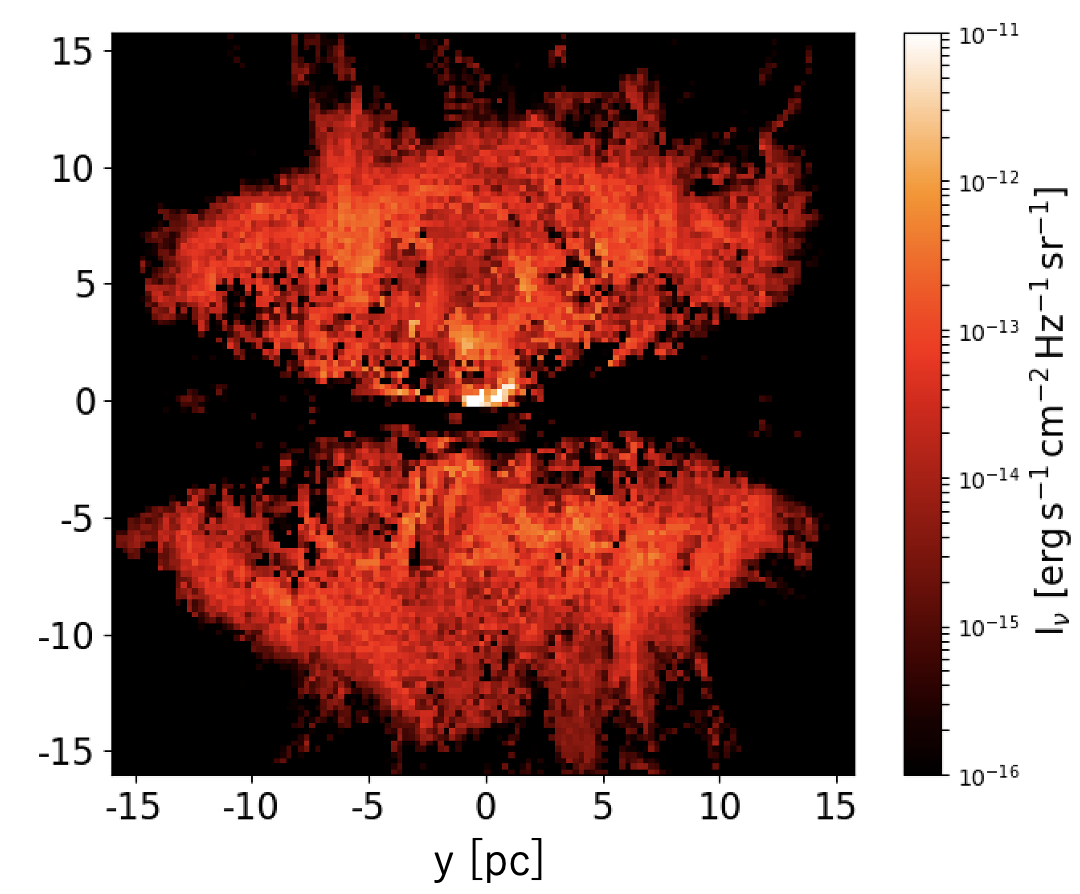}
    \caption{\label{fig:result2}Intensity map of thermal dust emission at the wavelength $4.7 \, \micro$ in the case we observe the torus at the inclination angle $\obs=77 \, ^{\circ}$ and azimuthal angle $\phi_\mathrm{obs}=0 \, ^{\circ}$.} 
\end{figure}

Figure \ref{fig:result2} shows the intensity map of thermal dust emission at a wavelength of $4.7\, \micron$ and at an inclination angle $\obs=77 \, ^{\circ}$.
The thermal emission from the hot dust in the central $1.5$-pc region is significant compared to the surrounding emission from the warm dust on a $10$-pc scale.
However, on the shadow area in the middle of Figure \ref{fig:result2}, the thermal emission is self-shielded by optically thick cold dust (the intensity maps of the dust continuum for various inclination angles are shown in Appendix \ref{ap:3}), and the area of self-shielded thermal emission from the hot dust increases as the inclination angle increases.


\begin{figure}[htb!]
    \centering
    \includegraphics[width=1.0\columnwidth]{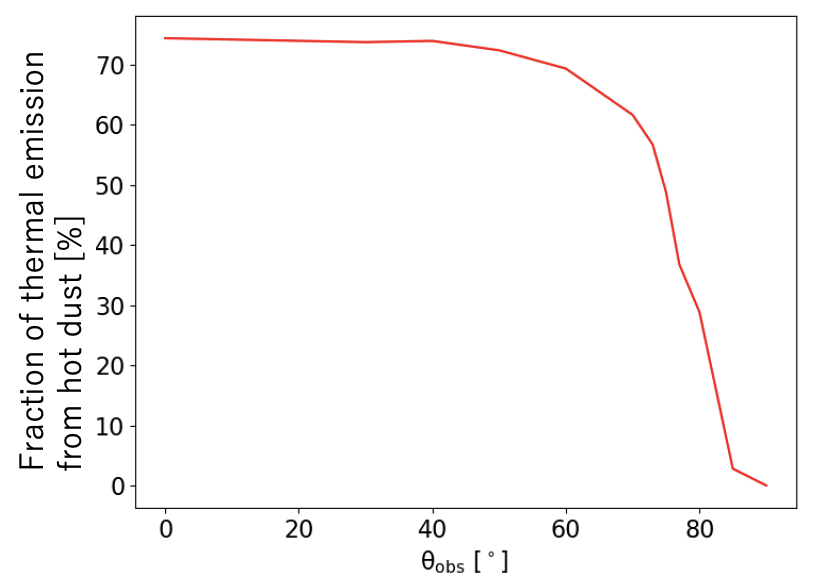}
    \caption{\label{fig:result3}Fraction of the thermal emission from the hot dust ($T_\mathrm{dust}>200$ K) in the innermost region to the total thermal emission captured within the field of view $32\times32 \,\mathrm{pc}^2$.} 
\end{figure}

To explore how much radiation from the hot dust contributes to $4.7\, \micron$ dust continuum, we investigate the fraction of the thermal emission from the hot dust in the innermost region to the total thermal emission captured within the field of view $32\times32 \,\mathrm{pc}^2$ at different inclination angles, as shown in Figure \ref{fig:result3}.
The fraction at the inclination angles $\obs=0 \, ^{\circ}-60 \, ^{\circ}$ is almost constant at $70$\%, 
whereas the fraction at the inclination angles $\obs=60 \, ^{\circ} - 90 \, ^{\circ}$ decreases from $65$\% to $0$\% steeply due to self-shielding of the cold dust.
Moreover, the fraction is higher than 30\% at the inclination angles $\obs \leq 80 \, ^{\circ}$, and thus, the thermal emission from the hot dust is effective as a background source to observe CO absorption at those inclination angles.

We note that the thermal emission from the hot dust in the innermost region is perfectly self-shielded by the cold dust at an inclination angle $\obs=90 \, ^{\circ}$.
Therefore, we expect that CO rovibrational absorption can be observed preferentially for smaller inclination angles.   



\subsection{Conditions for detecting CO absorption lines} \label{sec:result2}

Here, we show the results of the non-LTE line transfer calculations, and we investigate the conditions to observe CO rovibrational absorption lines against thermal emission from the hot dust in the near-infrared.
\begin{figure}[htb!]
    \centering
    \includegraphics[width=1.0\columnwidth]{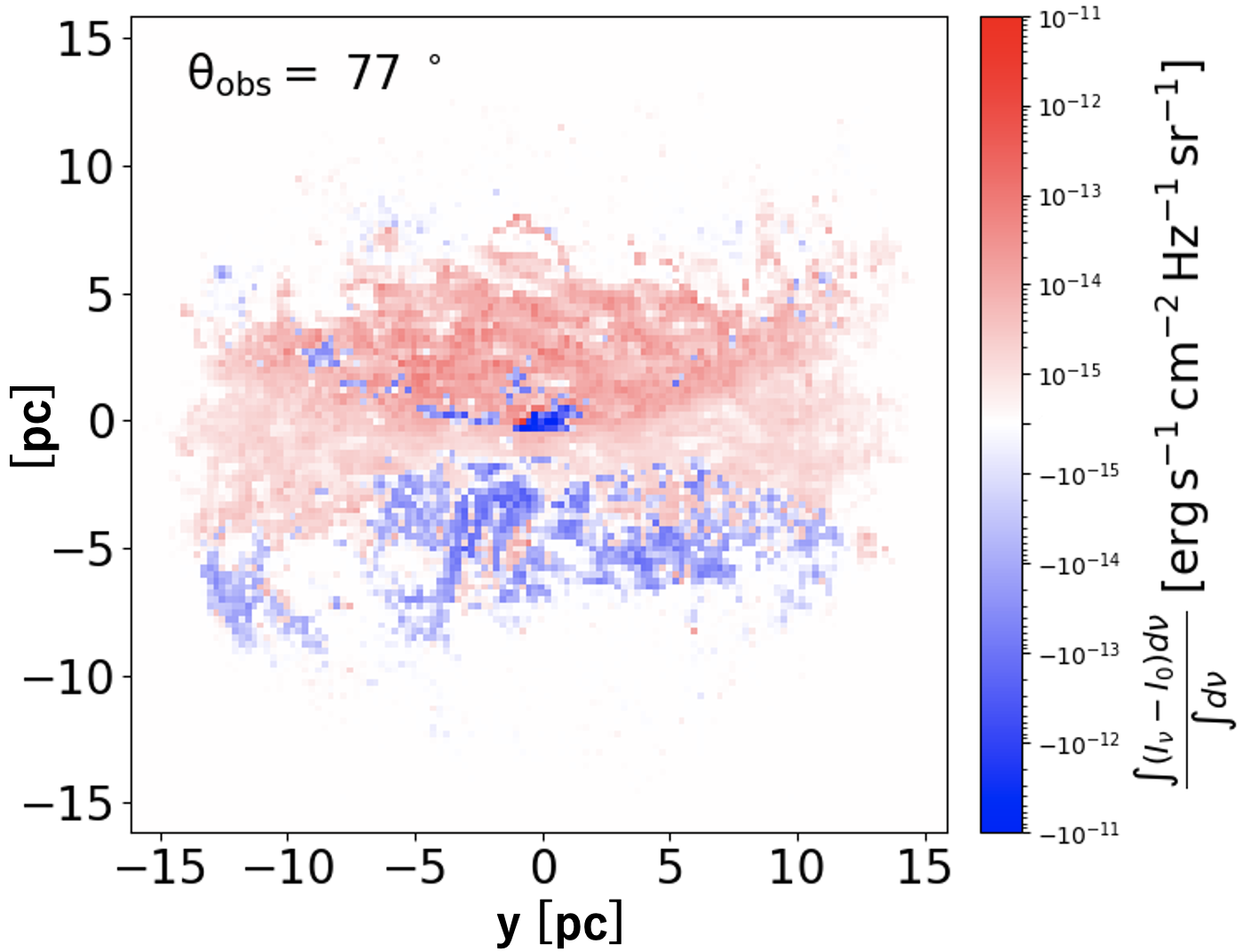}
    \caption{\label{fig:result4}A dust-continuum-subtracted intensity map around the wavelength corresponding to a rovibrational transition R(0) for the inclination angle $\obs=77 \, ^{\circ}$ and azimuthal angle $\phi_\mathrm{obs} =0 \, ^{\circ}$. 
    The color represents the line intensities, and  the reddish and the blueish colors show the intensities of the line in emission and absorption, respectively. 
    Here, the intensity is averaged over $-40$ to $40$ $\kms$.
    } 
\end{figure}
Figure \ref{fig:result4} shows a dust-continuum-subtracted intensity map around the wavelength corresponding to a rovibrational transition R(0).
The CO emission in the reddish color is visible within the entire torus, which is seen as a dark lane in Figure \ref{fig:result2}.
In the central region of Figure \ref{fig:result4}, the CO absorption in the blueish color is strongly seen against the thermal emission from the hot dust in the innermost region.
\begin{figure}[htb!]
    \centering
    \includegraphics[width=1.0\columnwidth]{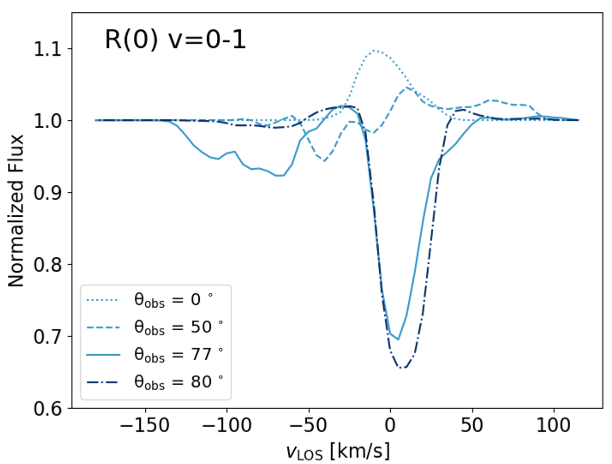}
    \caption{\label{fig:result5}Spectral line profiles of the CO rovibrational transition R(0) at various inclination angles $\obs$ and azimuthal angle $\phi_\mathrm{obs} = 0 \, ^{\circ}$. Here, the beam size and the spectral resolution are $32\times32 \,\mathrm{pc}^2$ and $5\, \kms$, respectively.} 
\end{figure}

To explore whether CO rovibrational absorption line can be detected for various inclination angles with the beam size of $32\times32 \,\mathrm{pc}^2$, i.e., the maximum area of the calculation box, we investigate the dependency of the CO rovibrational line profiles on various inclination angles\footnote{We calculate CO rovibrational line profiles at inclination angles $\obs = 10,\,20,\,30,\,40,\,50,\,60,\,70,\,73,\,75,\,77,\,80,$ and $90 \, ^{\circ}$.}.
Here, we integrate spectra of all grids in the field of view $32\times32 \,\mathrm{pc}^2$, and each integrated spectrum is normalized by the dust continuum.
Figure \ref{fig:result5} shows the spectral line profiles of the CO rovibrational transition R(0) at different inclination angles $\obs$ and the azimiuthal angle is fixed at $\phi_{obs}=0$.
At $\obs = 0 \, ^{\circ}$ , emission lines are observed, but absorption lines are not observed.  
Also, at the inclination angle $\obs = 50 \, ^{\circ}$, absorption lines are hardly observed.
Thus, at smaller inclination angles, such as the inclination angles $\obs = 0-50 \, ^{\circ}$, the column density $N_\mathrm{CO}$ on the line of sight is small, so CO absorption lines are not observed.
However, at the inclination angle $\obs = 77 \, ^{\circ}$, we observe the absorption line spectrum composed of multiple velocity components, including blue-shifted components (i.e., outflow components), a systemic component ($v_{\rm{LOS}}=0 \, \kms$), and a red-shifted component (i.e., inflow component).
At the inclination angle $\obs = 80 \, ^{\circ}$, only an absorption line of systemic component is observed. 
Therefore, CO absorption lines can be detected for inclination angles $\obs = 50-80 \, ^{\circ}$.

We also study the dependency of the CO absorption lines on the azimuthal angles; we examine whether CO absorption is detected at azimuthal angles from $\phi_\mathrm{obs}=-180 \, ^{\circ}$ to $180 \, ^{\circ}$ in $15 \, ^{\circ}$ increments.
This analysis roughly corresponds to examining the dependency in the time for a short period, since the torus is rotating, and the inner structures of the inhomogeneous torus does not change much with rotation for a short period.
We find that the range of inclination angles over which the systemic component of CO absorption is observed does not change for most azimuthal angles between $\obs = 50-80 \, ^{\circ}$.
On the other hand, we note that outflow and inflow components in the absorption profiles are detected only for some azimuthal angles; in the case with an inclination angle $\obs = 77 \, ^{\circ}$, outflow components are detected at azimuthal angles $\phi_\mathrm{obs}=-15-60 \, ^{\circ}$; inflow components are detected only at azimuthal angles $\phi_\mathrm{obs}=-30, \,0, \,120,$ and $135 \, ^{\circ}$.
The detection probabilities of the outflow and inflow components are $25 \, \%$ and $17 \, \%$, respectively.

Therefore, we conclude that CO rovibrational absorption lines can be detected in only the case with inclination angles $\obs = 50-80 \, ^{\circ}$, but the detectability of outflow and inflow components depends on the azimuthal angles.
(in Appendix \ref{ap:3}, we discuss differences in observational conditions between our results in the near-infrared wavelength with a previous theoretical study in the submillimeter wavelengths \citep{Uzuo2021}).



\subsection{Velocity components of CO rovibrational absorption lines} \label{sec:result3}

\begin{figure}[htb!]
    \centering
    \includegraphics[width=1.0\columnwidth]{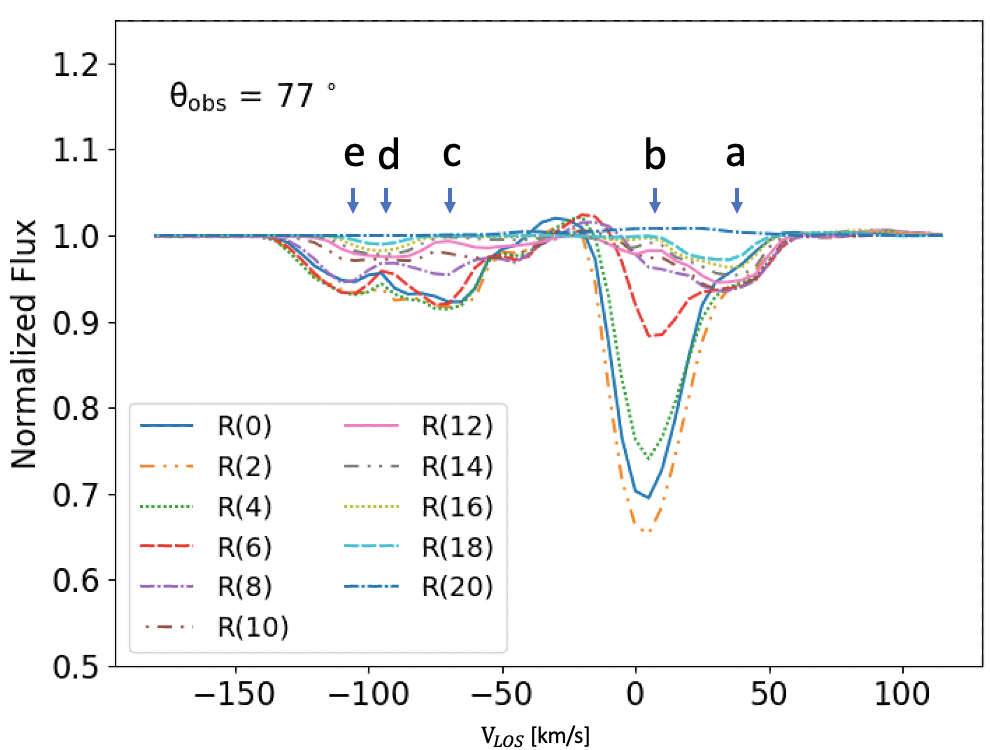}
    \caption{\label{fig:result6}Spectral line profiles of various rovibrational transitions R(J) at the inclination angle $\obs = 77 \, ^{\circ}$ and azimuthal angle $\phi_\mathrm{obs} = 0 \, ^{\circ}$.
    The symbols a, b, c, d, and e show the positions of the different velocity components in the absorption profiles, and properties of the velocity components are listed in table \ref{tab:3-1}. Here, the beam size and the spectral resolution are $32\times32 \,\mathrm{pc}^2$ and $5\, \kms$, respectively.} 
\end{figure}
To investigate the velocity components of CO rovibrational absorption lines, we focus on the case with an inclination angle $\obs = 77 \, ^{\circ}$ and an azimuthal angle $\phi_{obs}=0 \, ^{\circ}$, at which the CO rovibrational absorption spectrum includes several velocity components similar to the observational results \citep{Shirahata2013,Onishi2021}.
Figure \ref{fig:result6} shows the spectral line profiles of various rovibrational transitions R(0)-R(20).
We find five velocity components ($v_{\rm{LOS}}=+30,\, 0,\,-70,\, -95,$ and $-105\, \kms$) in the spectra.
Component (b) ($v_{\rm{LOS}}=0\, \kms$), which is the deepest absorption of all velocity components, shows the absorption lines with rotational levels up to R(8).
Also, components (c) and (e) ($v_{\rm{LOS}}=-70$ and $-105 \, \kms$) show the absorption lines with rotational levels up to R(8) and R(10), respectively.
Furthermore, components (a) and (d) ($v_{\rm{LOS}}=+30$ and $ -95\ \kms$) show the absorption lines with higher rotational levels up to R(18). 
Thus, we find that components (a) and (d) are excited to higher rotational levels compared to components (b), (c), and (e).

To explore physical values of the five velocity components, we estimate the observed column densities of CO rovibrational absorption lines of each velocity component and make the rotation diagram, i.e., the observed column densities expressed as a function of rotational level $J$.  
Here, observed column density is derived from averaging the column density on the line of sight in front of the hot dust emission over the entire field of view by weighting the intensity of continuum level of each grid.
Rotation diagram can be used to constrain the molecular gas physical conditions, such as the gas density and the excitation temperature.

\begin{deluxetable*}{cccccccc}
 \tablecaption{The physical values of the five velocity components of CO rovibrational absorption lines in the case with the inclination angle $\obs = 77 \, ^{\circ}$.
 }
 \label{tab:3-1}
 \tablewidth{0pt}
 \tablehead{
 \colhead{$\rm ID$} & \colhead{$v_{\rm{LOS}}$} & \colhead{Direction} & \colhead{$N_\mathrm{CO}$} &\colhead{$N_\mathrm{H_2}$} & \colhead{$T_\mathrm{ex}$ ($J=0-3$)} & \colhead{$T_\mathrm{ex}$ ($J=17-19$)} & \colhead{$r_\mathrm{abs}$}\\
 \colhead{} & \colhead{($\kms$)} & \colhead{} & \colhead{($\mathrm{cm^{-2}}$)} &
 \colhead{($\mathrm{cm^{-2}}$)} & \colhead{(K)} & \colhead{(K)}  & \colhead{(pc)} 
}
\decimalcolnumbers
\startdata
  (a)   & $+30 $ & inflow &   $8.10 \times 10^{17}$ & $2.96 \times 10^{21}$ & $186$ & $115$ & $1.5$ \\
  (b)   & $0 $ & systemic &   $9.42 \times 10^{16}$ & $3.25 \times 10^{20}$ & $24$ & $218$ & $1.0-7.5$\\
  (c)   & $-75 $ & outflow &  $5.06 \times 10^{16}$ & $1.88 \times 10^{20}$ & $33$ & $55$ & $1.0$\\
  (d)   & $-95 $ & outflow &  $2.23 \times 10^{18}$ & $8.15 \times 10^{21}$ & $380$ & $145$ & $0.75$\\
  (e)   & $-105 $ & outflow &   $4.46 \times 10^{16}$ & $6.89 \times 10^{18}$ & $65$ & $50$ &$1.0$\\
\enddata
\tablecomments{Column (1): ID of velocity components. Column (2): $v_{\rm{LOS}}$ is the line-of-sight velocity. Column (3): Classification of the velocity components. Columns (4) and (5): $N_\mathrm{CO}$ and $N_\mathrm{H_2}$ are the observed column densities of H$_2$ and CO, respectively, which are derived from averaging the column density on the line of sight in front of the hot dust emission over the entire field of view by weighting the intensity of continuum level of each grid. Column (6): $T_\mathrm{ex}$ is the excitation temperature. Column (7): $r_\mathrm{abs}$ is the distance of each velocity component from the AGN center.}
\end{deluxetable*}

\begin{figure*}[htb!]
    \vspace{-30pt}
    \centering
    \includegraphics[width=2.0\columnwidth]{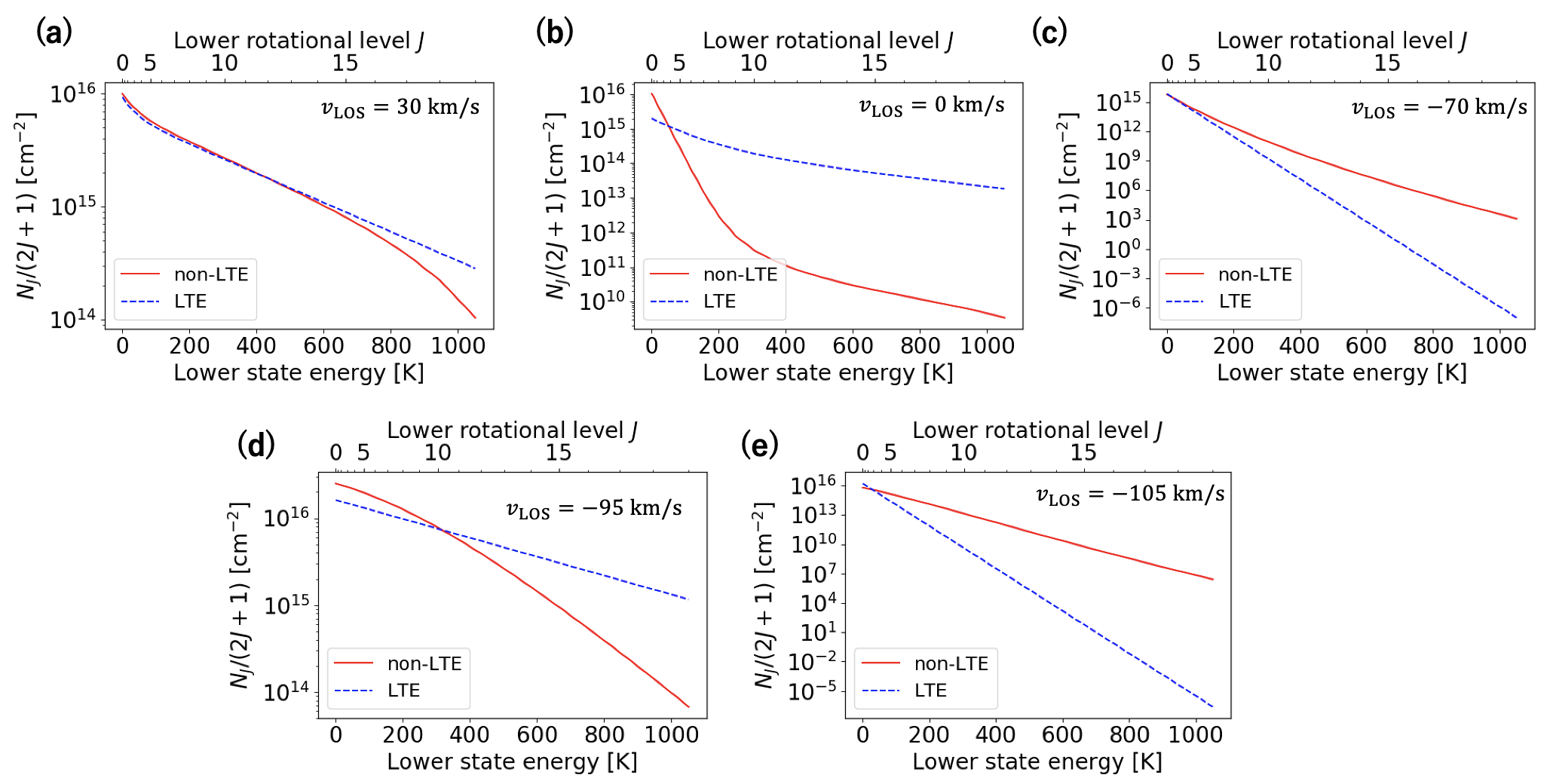}
    \caption{\label{fig:result12}Rotation diagrams of the five velocity components. The red solid and blue dashed lines represent the rotation diagram which is calculated by the non-LTE calculation and calculated using LTE assumption, respectively. When the main transition process is the radiational transition at each rotational level, the slope of the non-LTE calculation (non-LTE slope) is different from the slope of LTE assumption (LTE slope). Here, the column density $N_{J}$ is derived from averaging the column density on the line of sight in front of hot dust emission over the entire field of view by weighting the intensity of the continuum level of each grid.} 
\end{figure*}
Figure \ref{fig:result12} compares two rotation diagrams obtained from the non-LTE calculation and calculated using LTE assumption ($T_\mathrm{ex}=T_\mathrm{kin}$) for the five velocity components, which are summarized in Table \ref{tab:3-1}.
Components (b) and (e) show different slopes of rotation diagrams between the non-LTE and LTE.
That means that those components are absorbed by a low-density gas ($n_\mathrm{H_2} < 10^3 \, \mathrm{cm^{-3}}$, the critical density for $J=0-1$), and the main transition processes are radiational transitions.
In component (c), the non-LTE and LTE slopes coincide up to $J=3$, and the non-LTE slope gradually becomes gentle at higher rotational levels $J$ than the LTE slope.
This trend of the non-LTE implies that component (c) is absorbed by the gas with density $n_\mathrm{H_2} \sim 10^4 \, \mathrm{cm^{-3}}$ (the critical density for $J=2-3$), and the main transition process at only low $J$ ($J<3$) is collisional transition.
In both components (a) and (d), the non-LTE and LTE slopes coincide up to $J=13$ and $5$, respectively, and the non-LTE slopes gradually become steeper at higher rotational levels $J$ than the LTE slopes.
That means that components (a) and (d) are absorbed by high-density gas ($n_\mathrm{H_2} \sim 10^6 \ \mathrm{and} \ 10^5 \, \mathrm{cm^{-3}}$, the critical densities for $J=12-13$ and $J=4-5$, respectively), and the main transition processes up to $J=13$ and $5$ are collisional transitions.

Besides, we estimate the excitation temperatures for $J=0-3$ and $J=17-19$ of each velocity component by fitting the column densities of the non-LTE at the three rotational levels as
\begin{linenomath}
\begin{equation}
N_{J}/g_{J} \propto \mathrm{exp}(-h\nu/kT_\mathrm{ex}),
\label{eq:4.5}
\end{equation}
\end{linenomath}
where $g_J$ is the degeneracy of each rotational transition level $J$.
The excitation temperatures of components (b), (c) and (e) are $24$, $33$, and $65$ K at lower $J$ ($J=0-3$), respectively.
The excitation temperatures at lower $J$ ($J=0-3$) of components (a) and (d) reflect their kinetic temperature of $186$ and $380$ K, respectively, and the excitation temperatures gradually decrease to $115$ and $145$ K at higher $J$ ($J=17-19$), respectively.
Therefore, an inflow component (a) and an outflow component (d) are composed of highly dense and hot gas.
Physical parameters of the velocity components are summarized in Table \ref{tab:3-1}, and the origins of those components are discussed in Section \ref{sec:discuss2}.


\section{Discussion}\label{sec:discuss}
\subsection{Observations of CO rovibrational absorption lines} \label{sec:discuss1}
In Section \ref{sec:result1} and \ref{sec:result2}, we suggest that CO rovibrational absorption lines are detectable for the inclination angles $\obs = 50-80 \, ^{\circ}$. 
That means that CO absorption lines are more easily detected in type 2 AGNs than in type 1 AGNs.
In observations, \citet{2018PhDT.......191B} investigated CO rovibrational absorption bands in nearby Seyfert AGNs using low-resolution spectroscopic data of AKARI, and he found that the absorption bands were not detected in type 1 AGNs\footnote{
A type 1 AGN, Mrk $334$, in which the CO absorption is detected, is excluded from their type 1 samples because Mrk $334$ is originally classified into type 1.8 AGNs \citep{Veron2010A&A...518A..10V}.}, but they were detected in type 2 AGNs.
This result is consistent with our study that CO absorption can be detected in type 2 AGNs.

We note that we observe CO rovibrational absorption lines against the thermal emission from the hot dust in the central $1.5$-pc region, and the $1.5$-pc thermal emission is perfectly self-shielded by optically thick cold dust for the large inclination angles $\obs > 80 \, ^{\circ}$ (see Figures \ref{fig:result3} and \ref{fig:ap3}).
Thus, it is unnecessary that CO rovibrational absorption lines are detected in all type 2 AGNs, especially with the inclination angles $\obs > 80 \, ^{\circ}$.
In observations, \citet{Lutz2004} investigated CO rovibrational absorption lines in 31 Seyfert AGNs using spectroscopic data of ISO, and they found no significant detection of CO rovibrational absorption bands in all galaxies including type 2 AGNs (e.g., NGC1068).
To explain this result, we assume that self-shielding of dust is a reason why CO absorption lines are not always observed in type 2 AGNs.
Therefore, our result suggests that CO absorption lines can be observed at limited inclination angles $\obs \simeq 50-80 \, ^{\circ}$.





\subsection{Origins of the velocity components}\label{sec:discuss2}
\begin{figure*}[htb!]
    \centering
    \includegraphics[width=2.0\columnwidth]{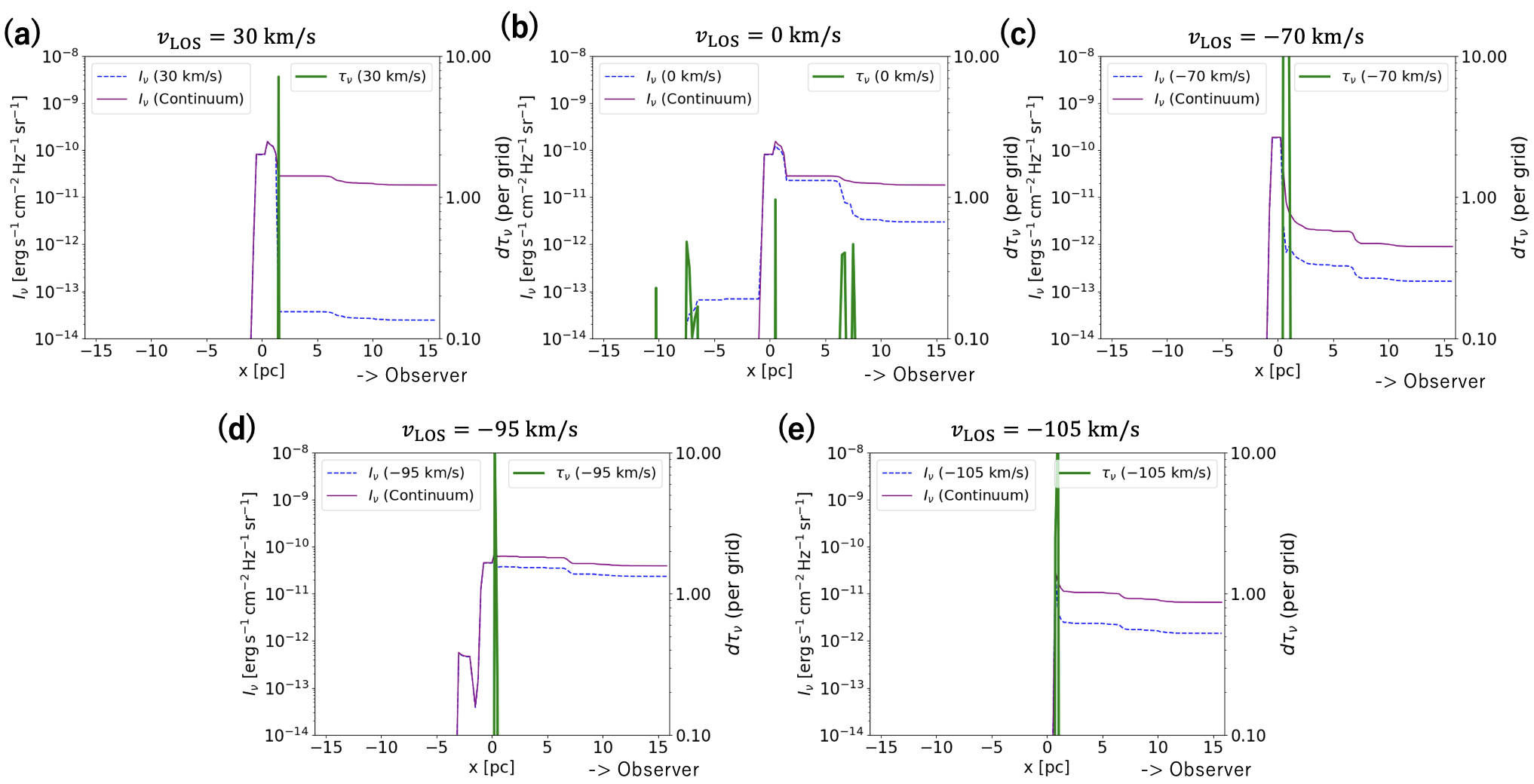}
    \caption{\label{fig:result7}Distributions of the intensity, optical depth, and continuum as a result of the radiative transfer from the far side to the near side. 
    The blue dashed and purple lines represent distributions of the intensity for each velocity component and of continuum level, respectively.
    The green line represents the optical depth per grid for each velocity component.
    When the optical depth (Green line) is high, the intensity for a velocity component (blue dashed line) decreases.
    } 
\end{figure*}

In Section \ref{sec:result3}, we observed CO rovibrational absorption lines including the systemic component with low excitation temperature ($T_\mathrm{ex}\sim$ a few tens K) and the outflow and inflow components with high excitation temperature ($T_\mathrm{ex}>100$ K).
This result is consistent with a recent near-infrared observation \citep{Onishi2021}, which shows CO rovibrational absorption lines including a systemic component with the excitation temperature $T_\mathrm{ex}=26 \pm 3$ K, outflow components with the excitation temperature $T_\mathrm{ex}=721 \pm 43$ and $T_\mathrm{ex}=\ 248 \pm 41$ K, and an inflow component with excitation temperature $T_\mathrm{ex}=1218 \pm 421$ K, in IRAS $08572$+$3915$.

Here, we investigate the origin of each velocity component of CO rovibrational absorption lines by tracing the intensity change along a line of sight for each velocity component.
Figure \ref{fig:result7} shows distributions of the intensity, optical depth, and continuum as a result of the radiative transfer from the far side to the near side.
Figure \ref{fig:result7} (a) shows the distributions of component (a) ($v_{\rm{LOS}}=+30\,\kms$), and the intensity $I_\nu$ (the blue dashed line) becomes lower than dust continuum level (the purple line) at $1.5$ pc from the center of the AGN ($r_\mathrm{abs} = 1.5$ pc) due to optically thick CO molecules.
That means that component (a) corresponds to accreting gas at $r_\mathrm{abs} = 1.5$ pc on the equatorial plane.
Similarly, in the outflow components (c), (d), and (e), line absorption occurs at $r_\mathrm{abs} = 1.0$, $0.75$, and $1.0$ pc, respectively.
These gases in the inner torus correspond to the outflowing gas driven by AGN radiation pressure from the center of the torus.
Therefore, components (a), (c), (d), and (e) originates from the inner-wall region of the torus.
In particular, component (d) is located at the innermost region of the torus and directly heated by the radiation from the accretion disk, and thus, its excitation temperature is the highest of all velocity components.
In the systemic component (b), line absorption occurs at $r_\mathrm{abs} = 1.0$, $6.25$, and $7.5$ pc.
For the inclination angle $\obs = 77 \, ^{\circ}$, the gas located at $r_\mathrm{abs} = 7.5$ pc corresponds to the gas at the height of $1.75$ pc from the equatorial plane. 

\begin{figure}[htb!]
    \centering
    \includegraphics[width=1.0\columnwidth]{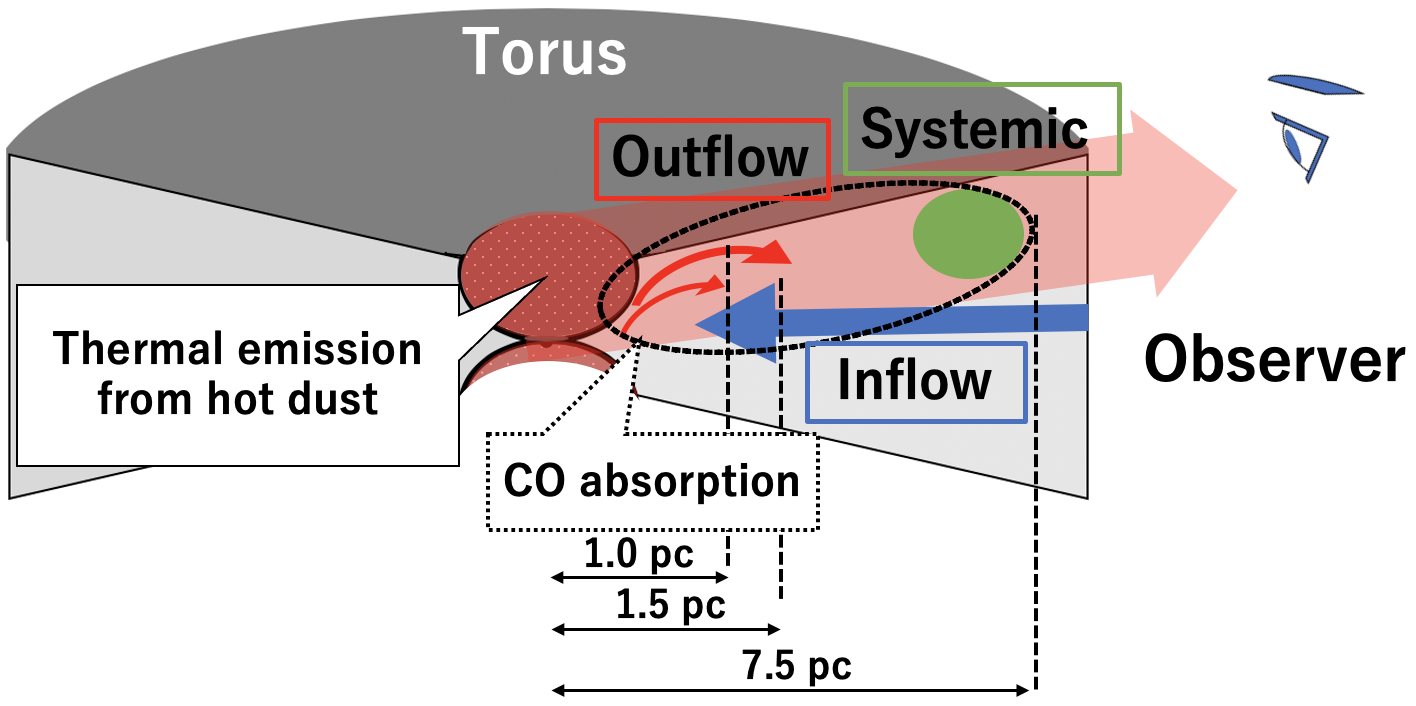}
    \caption{\label{fig:result8}A schematic picture of the observation for CO rovibrational absorption lines and the origin of the velocity components: outflow, systemic, and inflow.} 
\end{figure}

Figure \ref{fig:result8} shows schematically the locations of the inflow, outflow, and systemic components obtained from the absorption lines in this system.
The origins of all velocity components are summarized in Table \ref{tab:3-1}.
The inflow and outflow components correspond to the gas in the $1$-pc region near the dust sublimation layer, where a part of the inflow gas is converted to the outflow due to the radiation from the accretion disk as predicted by the radiation-driven fountain model \citep[e.g., ][]{Wada2016}.
Therefore, we can infer the “local” mass inflow or outflow rates using the velocity suggested by the absorption lines and gas density estimated by the critical densities, as shown in Section \ref{sec:result3}.
These rates are essential information on the fueling and feedback of AGNs, and it is still unclear even nearby AGNs.
Moreover, if both the inflow and outflow rates are confirmed, we can know the circulation of the mass in the torus, which is essential to sustain the torus thickness \citep{Wada2012}, and also estimate the fraction of the accreted gas to the central black hole using the luminosity of AGNs.

In addition to the above point, the excitation temperature at lower $J$ of the inflow and outflow components show the kinetic temperature higher than $100$ K as discussed in Section \ref{sec:result3}.
We confirm that inflow and outflow components detected as discussed in section \ref{sec:result2} also have high excitation temperatures at lower $J$ and originate in the inner $1$-pc region of the torus.
The temperature is higher than the typical temperature of molecular gas clouds in our Galaxy ($T\sim10$ K), and the gas is heated by UV and X-ray radiation in the inner region of the AGN torus (i.e., in PDR and XDR, respectively).
Therefore, such high excitation temperatures are probes that the gas observed by CO absorption lines originates in the inner region of the AGN torus.
Indeed, \citet[][]{2018PhDT.......191B} discovered gases with high excitation temperatures in tens of galaxies by CO rovibrational absorption lines.
Therefore, the inflow and outflow components are most likely to originate in the inner a-few-pc region of the AGN torus, depending on the dust sublimation radius ($r_\mathrm{abs} \propto L_\mathrm{AGN}^{0.5}$), and the CO absorption lines are likely to trace the gas circulation.

We should note that it is not clear whether all the AGN tori are similar to the Circinus’s torus; the detailed structures of the absorption lines can be different reflecting the inhomogeneous inner structures of the tori.
In fact, as \citet{Wada2015ApJ...812...82W} suggested, the structures of the radiation-driven fountain models change depending on the black hole mass and the AGN luminosity.
Therefore, we still need to compare further observations and hydrodynamic models based on the radiative transfer calculations in future studies. 

As shown above, these considerations suggest that we can obtain the velocities and temperatures of the inner a-few-pc region of AGN tori by observations of CO rovibrational absorption lines.
If we can observe multi-velocity components, the observations can be used as a probe of the gas circulation inside the AGN tori.
The feasibility of the detection of the multi-velocity components by observations is discussed in Section \ref{sec:discuss5}.


\subsection{Feasibility of future observations}
\label{sec:discuss5}
\begin{figure}[htb!]
    \centering
    \includegraphics[width=1.0\columnwidth]{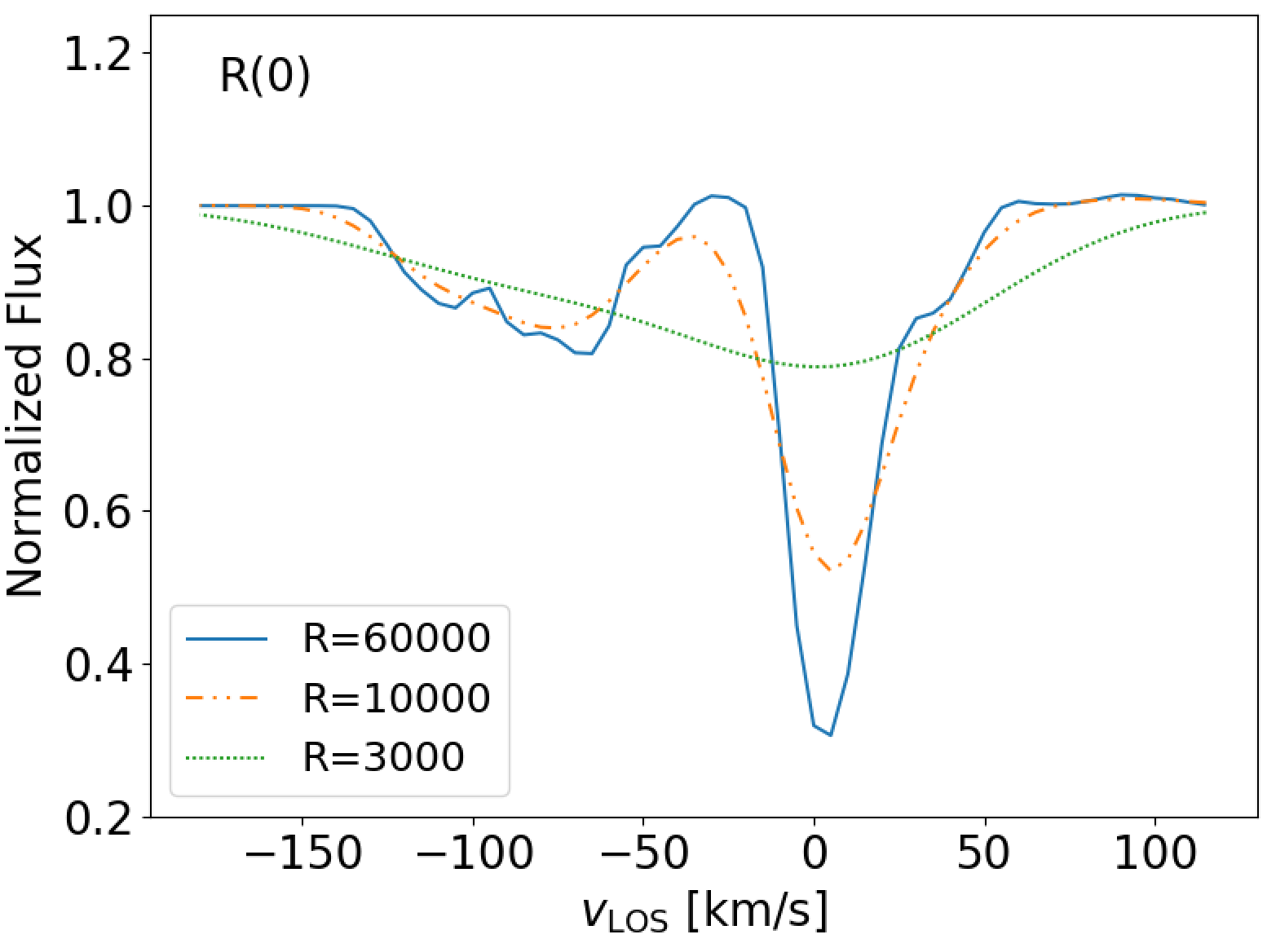}
    \caption{\label{fig:convolve} Blue solid line, orange dashed line, and green dotted line show CO rovibrational absorption profiles of R(0) at a inclination angle $\obs = 77 \, ^{\circ}$ with spectral resolutions of $R=60000$, $10000$, and $3000$ ($\Delta v = 5,\, 30,$ and $100$ $\kms$), respectively. The profile with $R=60000$ is the same as the profile of R(0) in Figure \ref{fig:result6}, and the profile with $10000$, and $3000$ are obtained from the convolution of the profile with $R=60000$ and Gaussian profiles of $R=3000$ and $10000$, respectively.} 
\end{figure}

To demonstrate the feasibility to detect CO rovibrational absorption lines with various telescopes, we investigate the CO rovibrational absorption profiles at the the representative angle of $\obs = 77 \, ^{\circ}$ and $\phi_\mathrm{obs} = 0 \, ^{\circ}$ with various spectral-resolution powers as shown in Figure \ref{fig:convolve}.
Here, the CO absorption profile of R(0) in Figure \ref{fig:result6} is convolved with Gaussian profiles of $R=10000$ and $3000$, which corresponds to the spectral resolutions of Infrared Camera and Spectrograph (IRCS)/Subaru telescope and Near Infrared Spectrograph (NIRspec)/James Webb Space Telescope (JWST), respectively.
The profile with $R=10000$ shows systemic and outflow components clearly, and hence we can decompose the absorption profile to systemic, and outflow components by fitting the profiles with a velocity-decomposition method as shown in \citet{Onishi2021}.
The profile with $R=3000$ shows only a systemic component extended to the negative velocity, and hence we would be able to decompose the absorption profile to systemic and outflow components by performing, e.g., double Gaussian fitting.
Therefore, we can derive velocities and temperatures of the systemic and outflow components with IRCS/Subaru and NIRspec/JWST.
The observation provides us with the presence of outflow in the gas circulation in the inner a-few-pc region of the AGN torus as discussed in Section \ref{sec:discuss2}, and these observations with several AGNs should be helpful to distinguish whether AGN tori are statically formed or dynamically formed, which is still under debate even nearby AGNs.

\subsection{Excitation mechanism: collision vs radiation} \label{sec:discuss4}

\begin{figure*}[htb!]
    \centering
    \includegraphics[width=2.0\columnwidth]{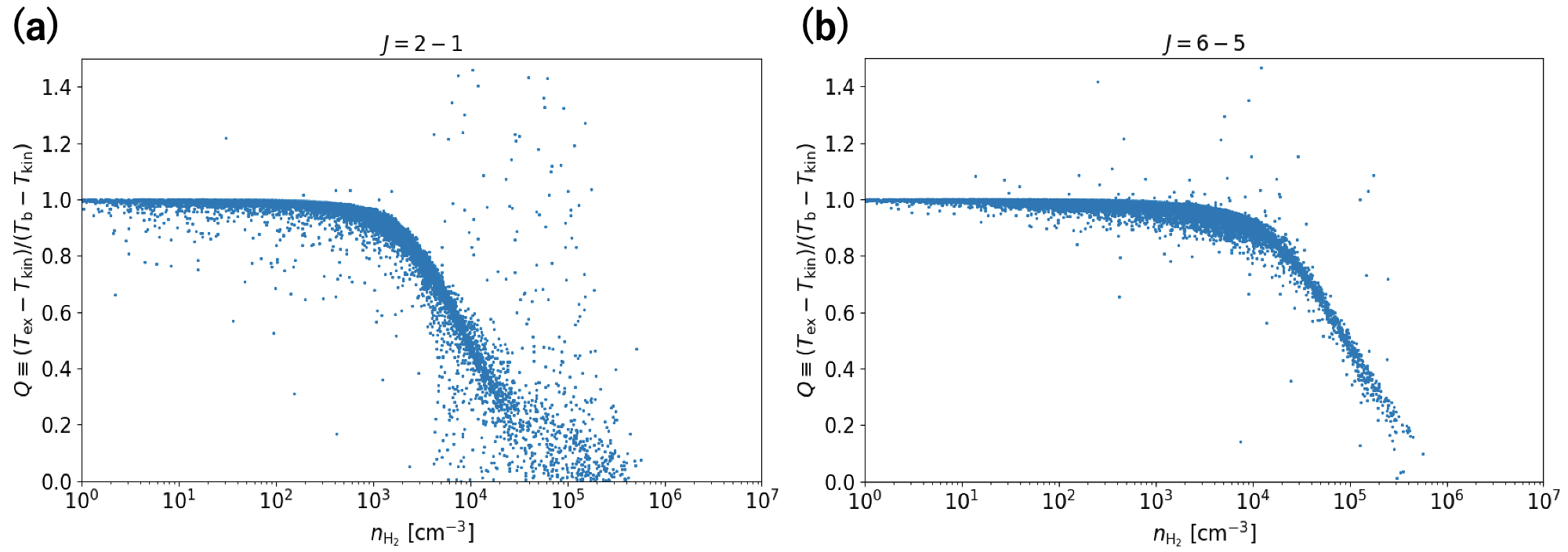}
    \caption{\label{fig:result10}(a) and (b) show dependency of $Q \equiv (T_\mathrm{ex}-T_\mathrm{kin})/(T_\mathrm{b}-T_\mathrm{kin})$ on the molecular hydrogen density of each grid with the rotational transition $J=2-1$ and $J=6-5$, respectively. The girds are taken from the slice of the calculation box with $\left|y\right|<2.5$ pc in the calculation box. Here, we exclude grids with $\left|T_\mathrm{b}-T_\mathrm{kin}\right|<5$ K because it is difficult to determine whether radiational excitation or collisional excitation is effective on those grids.
} 
\end{figure*}
\begin{figure*}[htb!]
    \centering
    \includegraphics[width=2.0\columnwidth]{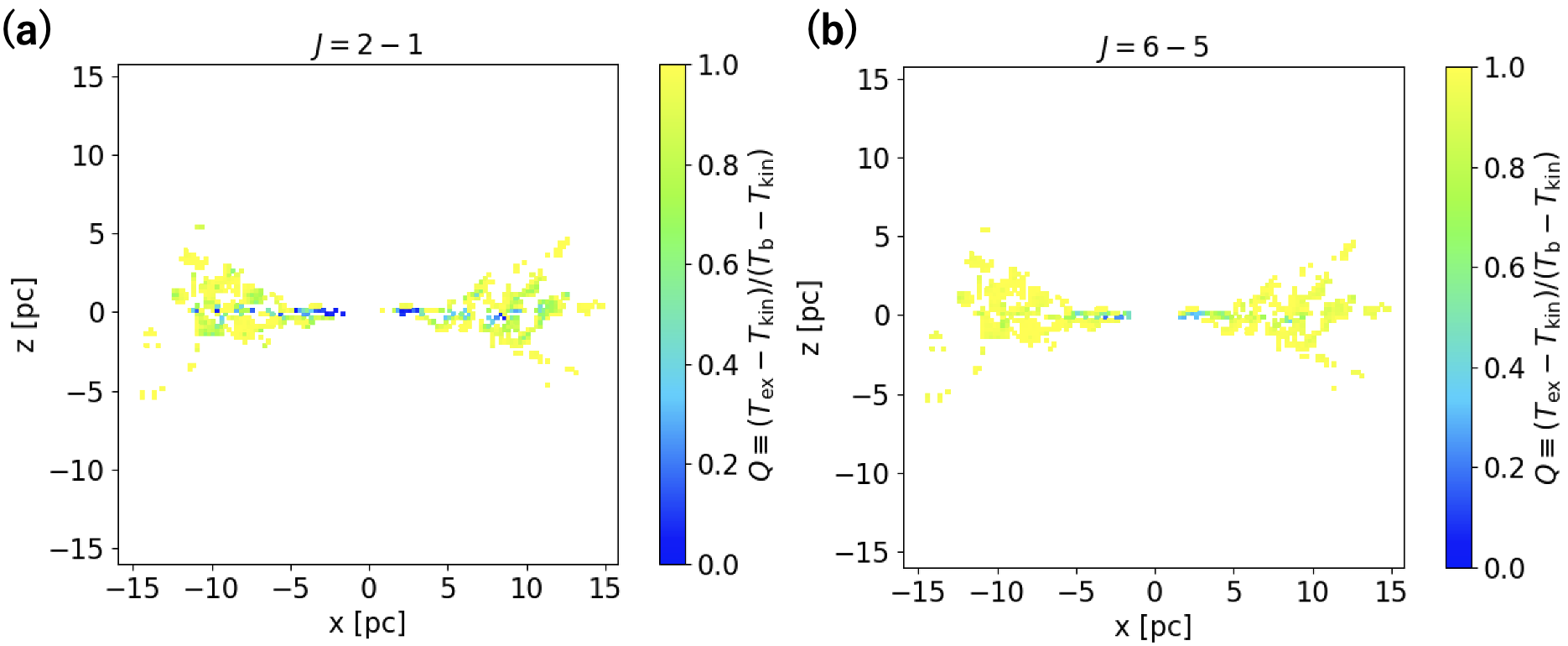}
    \caption{\label{fig:result11}(a) and (b) show maps of $Q \equiv (T_\mathrm{ex}-T_\mathrm{kin})/(T_\mathrm{b}-T_\mathrm{kin})$ with the rotational transition $J=2-1$ and $J=6-5$ on the x-z plane, respectively. The level population at a rotational transition is determined by radiational excitation ($Q\sim1$) or by collisional excitation ($Q\sim0$). We note that only the regions with $n_\mathrm{CO}>0.01$ cm$^{-3}$ are shown here.} 
\end{figure*}
In Section \ref{sec:result3}, Figure \ref{fig:result12} shows different shapes of the rotation diagrams between the non-LTE and the LTE for all velocity components, and it indicates that the observed level populations are not in the state of LTE.
Indeed, level populations depend on several factors: the gas density, kinetic temperature, and mean intensity of local radiation fields, and thus, what determines the excited states of CO molecules in molecular tori is always a matter of debate in observations \citep{Shirahata2013,Baba2018,Onishi2021}.
In this section, we explore which excitation mechanism of CO molecules at the rotational levels is dominant in the entire molecular torus, i.e., collisional excitation or radiational excitation (effects of rovibrational transition on rotational levels of the rovibrational level $v=0$ and the effects of the dust radiation fields are investigated in Appendix \ref{ap:1}).

To explore the excitation mechanism, we compare the excitation temperature with the kinetic temperature and the brightness temperature of the mean intensity in each grid.
Generally, in the case that collisional transitions are dominant to determine the level population, excitation temperature is equal to the kinetic temperature, and in the case that radiational transitions are dominant, excitation temperature is equal to the brightness temperature of mean intensity.
The excitation temperature between two rotational levels in the vibrational level $v=0$ is defined as
\begin{linenomath}
\begin{equation}
T_\mathrm{ex} \equiv \frac{h \nu}{k \log ((n_{J+1}/g_{J+1})/(n_{J}/g_{J}))}.
\label{eq:4.1}
\end{equation}
\end{linenomath}
The brightness temperature of the mean intensity in each grid is defined as
\begin{linenomath}
\begin{equation}
T_\mathrm{b} \equiv \frac{h \nu}{k \log(\frac{2h\nu^3}{c^2 J_{\nu}}+1)},
\label{eq:4.2}
\end{equation}
\end{linenomath}
where $J_{\nu}$ is the mean intensity in each grid.
To compare those temperatures in the entire torus, we define a parameter $Q$ as
\begin{linenomath}
\begin{equation}
Q \equiv \frac{T_\mathrm{ex}-T_\mathrm{kin}}{T_\mathrm{b}-T_\mathrm{kin}}.
\label{eq:4.3}
\end{equation}
\end{linenomath}
Figure \ref{fig:result10} shows the dependency of $Q$ on the molecular hydrogen density of each grid for the rotational transitions $J=2-1$ and $J=6-5$.
In Figure \ref{fig:result10} (a), if the density is lower than the critical density $10^3 \, \rm{cm^{-3}}$, $Q$ is equal to $1$.
In this density range, radiational transitions dominate the level transitions of CO molecules, and the excitation temperature is close to the brightness temperature.
If the density is larger than the critical density $10^3 \, \rm{cm^{-3}}$, $Q$ decreases.
In this density range, collisional excitation becomes more effective compared to radiative excitation to determine the level population, and the level population is determined by the balance of spontaneous emission and collisional transitions.
The excitation temperature is given by an analytic solution for a two-rotational-level system,
\begin{linenomath}
\begin{equation}
T_\mathrm{ex} = \frac{T_\mathrm{kin} \ E_{ul}}{k\, T_\mathrm{kin} \log (1+n_\mathrm{cr}/n_\mathrm{H_2})+E_{ul}}   \ \ (n_\mathrm{H_2}>n_\mathrm{cr}),
\label{eq:4.4}
\end{equation}
\end{linenomath}
where $E_{ul}$ is an energy gap between the upper and the lower levels.
Especially, if the density is significantly higher than the critical density (i.e., LTE case), the excitation temperature is equal to the kinetic temperature and $Q$ becomes $0$.
Figure \ref{fig:result10} (b) shows the dependency of $Q$ for the rotational transition $J=6-5$, and the $Q$ dependency is similar to that of $J=2-1$ though the critical density is $10^5 \, \rm{cm^{-3}}$.
Therefore, we can determine whether the excited states of the CO molecule are determined by radiative ($Q=1$) or collisional excitation ($Q=0$) by comparing the temperatures using $Q$.

Figure \ref{fig:result11} (a) shows the map of $Q$ for the rotational transition $J=2-1$ on the x-z plane.
In the grids on the equatorial plane, $Q$ is close to $0$, and thus, the rotational transition $J=2-1$ is determined mainly by collisional excitation.
However, in the region away from the equatorial plane, $Q$ is close to $1$, and hence, the rotational transition of $J=2-1$ is determined by radiational excitation.
Figure \ref{fig:result11} (b) shows the map of $Q$ for the rotational transition $J=6-5$.
Contrary to Figure \ref{fig:result11} (a), $Q$ is larger than $0$ in most grids, even on the equatorial plane.
That means that radiational transition can be effective at the higher rotational levels $J\geq5$ on the equatorial plane.

These results suggest that it is not the thermal equilibrium case in the entire molecular torus, and radiational excitation is effective in the torus, although collisional excitation is effective at lower rotational levels $J<5$ on the equatorial plane.


\section{Conclusions} \label{sec:concl}
The radiation-driven fountain model \citep{Wada2016} suggested a picture that AGN tori are naturally formed by gas circulation around the AGN, and this picture is consistent with multi-wavelength observations of low luminosity AGNs, such as the Circinus galaxy \citep{Wada2018a,Izumi2018,Wada2018b, Buchner2021, Ogawa2021}.
However, these previous studies have not directly verified the gas circulation inside the torus \citep[see also][]{Uzuo2021}.
In this paper, we propose that CO rovibrational absorption lines can be used to infer the internal motion of the molecular tori and focus on the following four questions: (1) what is the source of continuum radiation at the back of the CO absorption gas in near-infrared observations, especially at the wavelength $4.7\, \micron$?; (2) under what conditions are CO absorption lines detected?; (3) what is the origin of the gas traced by the CO rovibrational absorption lines?; (4) what determines the excited states of the observed CO molecules in molecular tori: collision or radiation?

To answer these questions, we performed two types of 3D radiative transfer calculations based on a snapshot of the radiation-driven fountain model. 
Firstly, to investigate the background source for CO rovibrational absorption lines at the wavelength $4.7\, \micron$, we performed 3D dust radiative transfer calculations.
Secondly, to explore conditions for detecting CO rovibrational absorption lines and the origin of each velocity component in the absorption lines, we performed 3D non-LTE line radiative transfer calculations.

The results are summarized as follows:
\begin{enumerate}
    \item
    Cold dust at a few ten K exists near the equatorial plane of the AGN torus, whereas warm dust at $100 - 200 \, \mathrm{K}$ exists in the cone-shaped inner-wall region of the torus.
    Hot dust at a temperature higher than $200 \, \mathrm{K}$ is concentrated in the inner $1.5$-pc region of the torus.
    The hot dust radiates effectively at the wavelength $4.7\, \micron$, as shown in Figures \ref{fig:result2} and Appendix \ref{fig:ap3}.
    \item
    We estimate the fraction of the thermal emission from the hot dust in the inner $1.5$-pc region to the total thermal emission captured within the field of view $32\times32 \,\mathrm{pc}^2$, as shown in Figure \ref{fig:result3}.
    The fraction decreases due to self-shielding of the cold dust as the inclination angle increase. Moreover, the fraction at the inclination angles $\obs \leq 80 \,^{\circ}$ is still higher than $30\%$.
    Thus, the radiation from the hot dust is effective as the background source of CO rovibrational absorption at the inclination angles $\obs\leq 80 \, ^{\circ}$.
    \item
    We investigate the conditions for detecting CO rovibrational absorption lines against the thermal emission of hot dust in the inner $1.5$-pc region of the torus. We find that CO rovibrational absorption lines can be detected in the case with the inclination angles $\obs = 50-80 \, ^{\circ}$ as shown in Figure \ref{fig:result5}. 
    \item
    In Figure \ref{fig:result6}, we find five velocity components ($v_{\rm{LOS}}=+30,\, 0,\,-70,\, -95,$ and $-105 \, \kms$) in the CO rovibrational spectra at the inclination angle $\obs = 77 \, ^{\circ}$ and the azimuthal angle $\phi_{\mathrm{obs}} = 0 \, ^{\circ}$. The excitation temperature of components (b), (c), and (e) are $24$, $33$, and $65$ K at lower $J$ ($J=0-3$), respectively. The excitation temperature at lower $J$ ($J=0-3$) of components (a) and (d) reflect their kinetic temperature of $186$ and $380$ K, respectively, and the excitation temperature gradually decrease to $115$ and $145$ K at higher $J$ ($J=17-19$), respectively.
    This result that the outflow and inflow components trace the gas with high excitation temperature ($T_\mathrm{ex}>100$ K) is consistent with a near-infrared observation result \citep{Onishi2021}.
    \item
    We investigate the origins of the five velocity components by tracing the intensity change along a line of sight (see Figure \ref{fig:result7}) and found their locations (see Figure \ref{fig:result8}).
    The inflow component (a) originates from accreting gas on the equatorial plane at $r_\mathrm{abs}=1.5$ pc from the AGN center.
    The outflow components (c), (d), and (e) originate from the outflowing gas at about $r_\mathrm{abs} \sim 1.0$ pc.
    The systemic component (b) originates from the gas at $r_\mathrm{abs}=1.0-7.5$ pc.

    \item
   We have introduced a parameter $Q$, which is defined as $Q \equiv (T_\mathrm{ex}-T_\mathrm{kin})/(T_\mathrm{b}-T_\mathrm{kin})$ (see Equation \ref{eq:4.3}), to indicate whether the level population of each energy state is determined mainly by collisional excitation ($Q\sim0$) or radiational excitation ($Q\sim1$).
    Using parameter $Q$, we find that radiational excitation is effective in the entire molecular torus, as shown in Figure \ref{fig:result10}, although collisional excitation is effective at lower rotational levels $J<5$ on the equatorial plane.
\end{enumerate}

These results suggest that CO rovibrational absorption lines can provide us with velocities and kinetic temperatures of inflows and outflows in the inner a-few-pc of tori.
The observations can probe the gas circulation inside the AGN tori.
We note that there are still only a few observations with decomposed velocity components in CO rovibrational absorption lines \citep{Onishi2021}, although there are dozens of galaxies in which CO rovibrational bands are detected with low-dispersion spectroscopy of AKARI and Spitzer \citep[][]{2018PhDT.......191B}.
Therefore, we suggest exploring the CO rovibrational absorption lines in those galaxies using the Subaru telescope or JWST, as discussed in Section \ref{sec:discuss5}.
\begin{linenomath}
\begin{acknowledgments}
The authors thank the anonymous referee for fruitful comments and suggestions to improve this paper.
We performed numerical calculations on the XC50 at the Center for Computational Astrophysics in the National Astronomical Observatory of Japan.
This work was supported by JSPS KAKENHI Grant Number 21H04496 (K.W. and T.N.), JP19J00892 (S.B.), and JP19J21010 (S.O.).
\end{acknowledgments}
\end{linenomath}
\appendix
\section{Intensity maps of dust continuum for the various inclination angles}\label{ap:3}
\begin{figure*}[htb!]
    \centering
    \includegraphics[width=2.0\columnwidth]{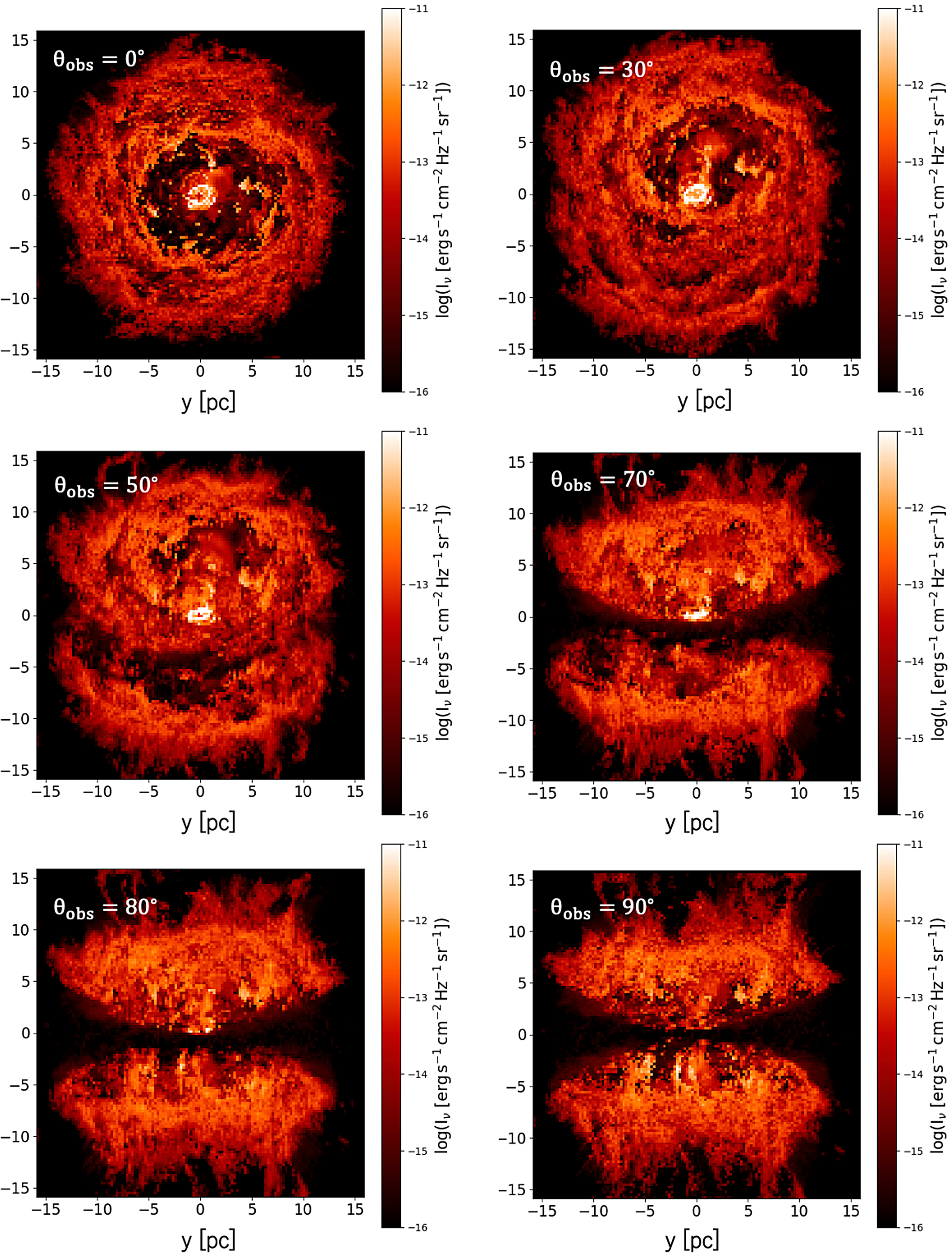}
    \label{fig:ap3}
    \caption{Intensity maps of dust continuum at the wavelength $4.7 \, \micro$ for the various inclination angles $\obs = 0, \, 30, \, 50, \,70,\, 80,$ and $90 \, ^{\circ}$} 
\end{figure*}

The intensity maps of dust continuum at the wavelength $4.7 \, \micro$ for the various inclination angles are shown in Figure \ref{fig:ap3}.
At the inclination angle  $\obs = 0 \, ^{\circ}$, hot dust radiates on the central $1.5$-pc ring, which is significant compared to the surrounding emission from the warm dust on a $10$-pc scale.
For the large inclination angles, the thermal emission from the hot dust is self-shielded by optically thick cold dust in the outer region of the torus.
The central $1.5$-pc ring may correspond to a ring-like continuum in the M band image observed by the Multi AperTure mid-Infrared SpectroScopic Experiment (MATISSE) of The Very Large Telescope Interferometer (VLTI) \citep[see Figure 1 of ][]{MATISSE2022Natur.602..403G}.

\section{Effects of the dust radiation field on CO rotational level population via rovibrational transitions}\label{ap:1}
Section \ref{sec:discuss2} suggests that radiational excitation is effective in molecular tori.
Here, we discuss whether radiational excitation due to dust radiation fields is effective in molecular tori by comparing three radiative transfer models described below.
\begin{description}
    \item[ROT (i.e., rotation)] \mbox{}\\
    This model includes only rotational transitions, and it considers only the radiation field of Cosmic Microwave Background (CMB), according to \citet{Wada2018a}.
    \item[ROTWDUST (i.e., rotation with dust)] \mbox{}\\
    This model includes only rotational transitions, and it considers not only the radiation field of CMB but also that of dust, according to \citet{Uzuo2021}
    \item[ROVIBWDUST (i.e., rovibration with dust)] \mbox{}\\
    This model includes both rotational transitions and rovibrational transitions, and it considers not only the radiation field of CMB but also that of dust.
\end{description}

\begin{figure*}[htb!]
    \centering
    \includegraphics[width=2.0\columnwidth]{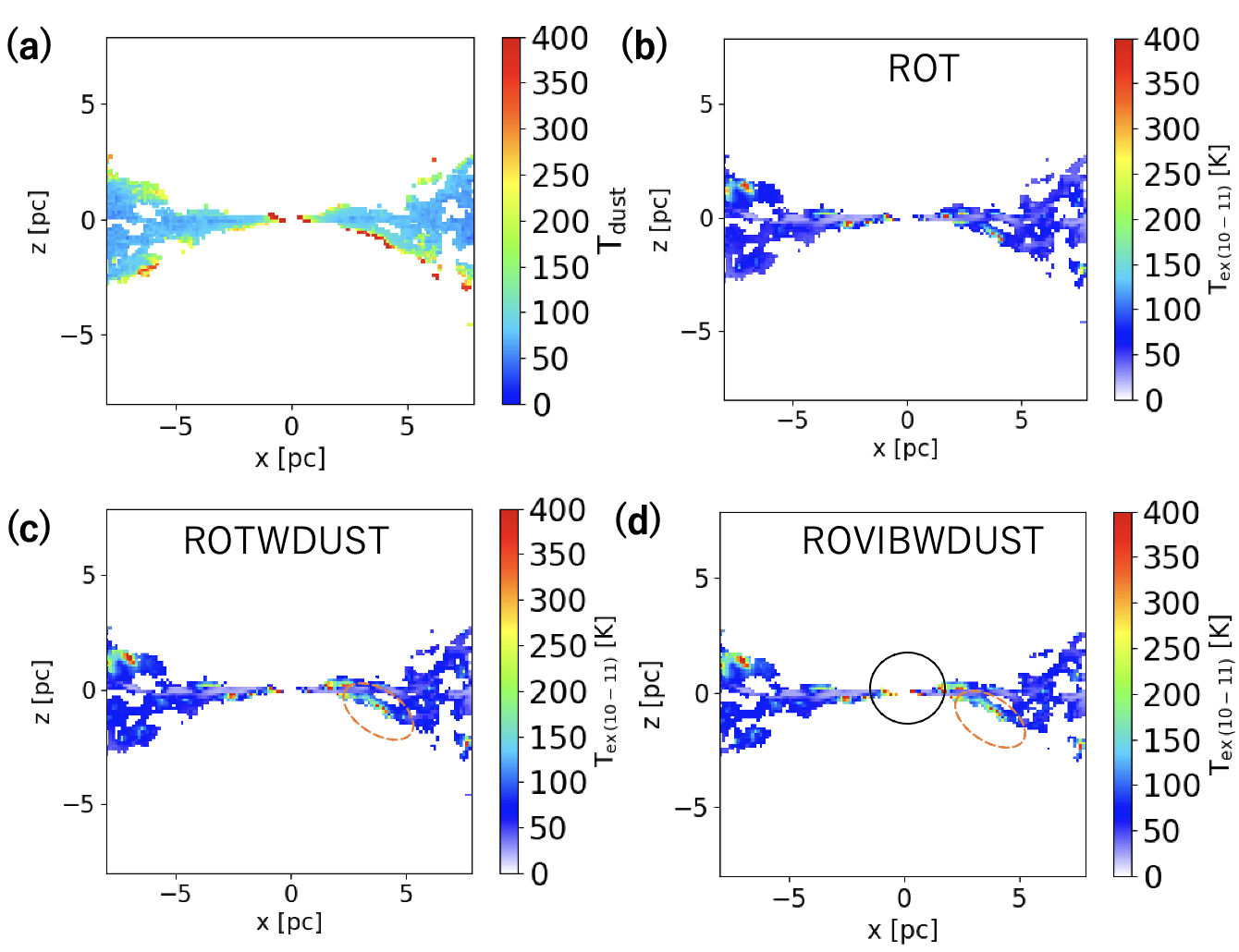}
    \caption{\label{fig:app1}(a) shows dust temperature map of the $0.1\, \micron$ graphite dust grains in the 16-pc box. (b), (c), and (d) show maps of the excitation temperatures between the two rotational levels $J=10-11$. The orange dashed circle represents the region where CO molecules are excited radiationally between the rotational levels. The black dashed circle represents the region where CO molecules are excited radiationally between the rotational levels via rovibrational transitions. We note that only regions with $n_\mathrm{CO}>0.01$ cm$^{-3}$ are shown, to focus on CO molecular gas properties.} 
\end{figure*}

We calculate those radiative transfer calculations based on the central 16-pc region of the calculation box with twice higher spatial resolution ($0.125$ pc) not to underestimate dust temperatures and the radiation fields due to the low resolution.
Figure \ref{fig:app1} (a) shows the dust temperature map on the x-z plane in the $16$-pc box, and there is hot dust at a temperature higher than $400$ K in the inner-wall region of the torus.
Figures \ref{fig:app1} (b), \ref{fig:app1} (c), and \ref{fig:app1} (d) show excitation temperature maps of the three models, and there is no obvious difference in the excitation temperatures in the torus except the inner-wall region.
Comparing Figures \ref{fig:app1} (c)  and \ref{fig:app1} (d) with Figure \ref{fig:app1} (b), we observed that the excitation temperatures in the inner-wall region of the ROTDUST and ROVIBWDUST are higher than the ROT (see the orange dashed circle).
That means that CO molecules in the inner-wall region are radiationally excited by dust radiation.
In the ROVIBWDUST, the excitation temperatures in the innermost region of the torus are significantly higher than those of the other models (see the black dashed circle).
This indicates that CO molecules are excited by dust radiation via rovibrational transitions.
In the innermost region, dust temperature is high ($T_\mathrm{dust}>400$ K) and optical depth at ${4.7 \, \micro}$ per grid is high.
Thus, the hot dust radiates significantly at the wavelength ${4.7 \, \micro}$ and excite CO molecules to the vibrational level $v=1$ ($B_{4.7 \, \micro}$($400$ K) $B_{J,J^{'}} \sim 10^{-2}$ (1/s)).
The vibrationally excited CO molecules immediately transit to the vibrational level $v=0$ due to spontaneous emission ($A_{J^{'},J} > 1$ (1/s)), and the high frequent rovibrational transitions affect determining rotational levels of the lower rovibrational level $v=0$.

\begin{figure}[htb!]
    \centering
    \includegraphics[width=1.0\columnwidth]{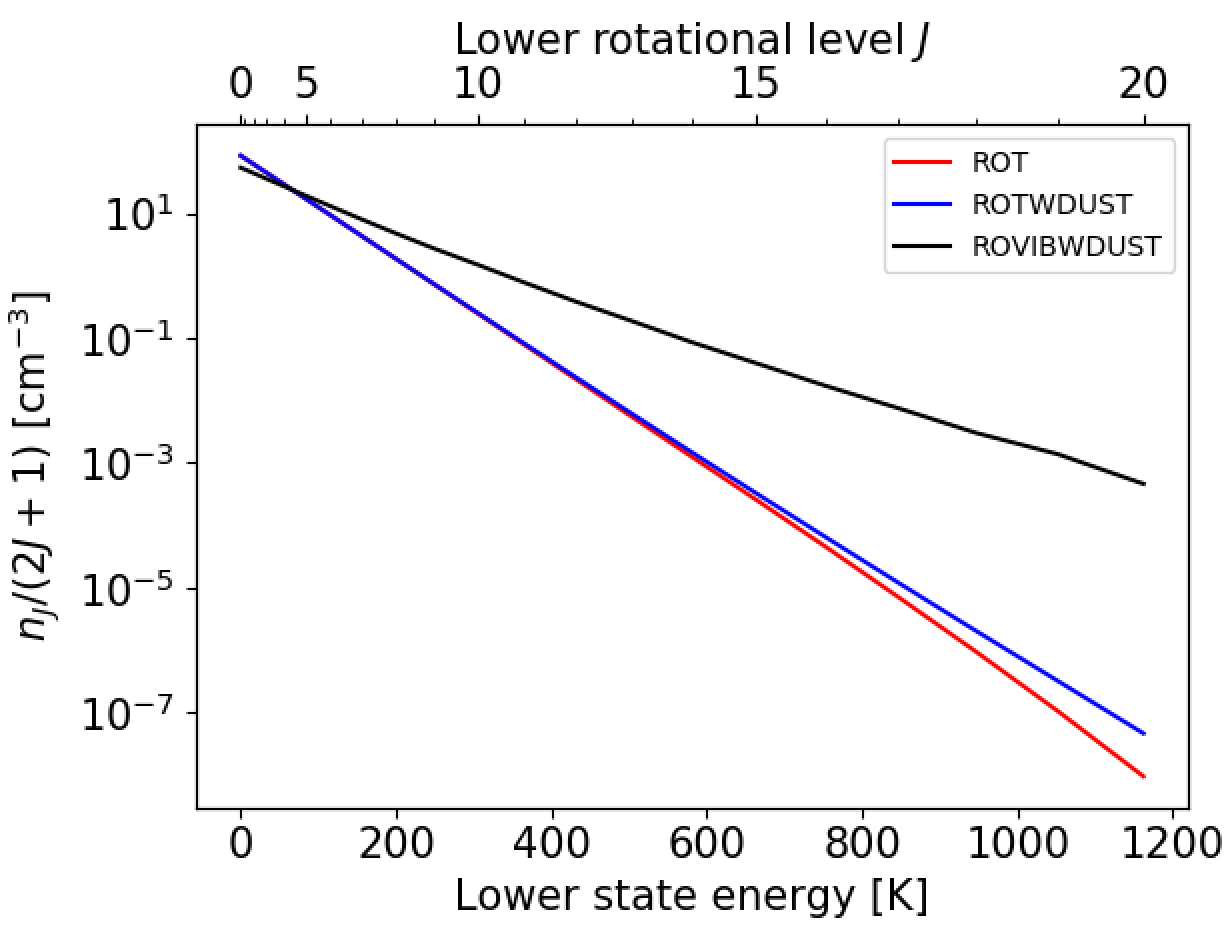}
    \caption{\label{fig:app3} Rotational diagrams of a grid at (x, y, z) = ($0.5$ pc, $0.125$ pc, $0.125$ pc) in the inner $1$-pc region of the torus. The red, blue, and black lines represent the rotational diagrams of ROT, ROTWDUST, and ROVIBWDUST models, respectively.} 
\end{figure}
Figure \ref{fig:app3} shows a rotational diagram of a grid in the inner $1$-pc region of the torus with the three models.
We note that the rotational diagram is the number density of CO molecules expressed as a function of the lower rotational levels $J$.
CO molecules in the ROVIBWDUST are excited up to higher rotational levels $J$ compared to the other models, though those in the ROTWDUST are a little excited at high $J$ ($J\geq15$) than in the ROT.
The difference between the ROVIBWDUST and the ROTWDUST is due to the dust optical depth in the near-infrared wavelength (rovibrational transitions: $\lambda \sim 4.7 \, \micro$) being significantly larger than those at the submillimeter wavelength (rotational transitions: $\lambda \sim 2600/(J+1) \, \micro$), as described in the following equations,
\begin{linenomath}
\begin{align}
& \tau_{ 4.7 \micro} \sim 1 \,  (\frac{dL}{0.125\, (\mathrm{pc})})\, (\frac{\rho_\mathrm{dust} }{10^{-20}\, (\mathrm{g/cm^{3}})}),
\label{eq:app:2}\\
& \tau_{J} \sim 2 \times10^{-3}  (J+1)^2 \,(\frac{dL}{0.125\, (\mathrm{pc})}) \,  (\frac{\rho_\mathrm{dust} }{10^{-20}\, (\mathrm{g/cm^{3}})}),
\label{eq:app:3}
\end{align}
\end{linenomath}
where $\tau_{ 4.7 \, \micro}$ and $\tau_{J}$ are optical depths of dust at the wavelengths $\lambda \sim 4.7$ and $2600/(J+1) \, \micro$, respectively.
Equation (\ref{eq:app:3}) assumes the dust opacity $\kappa_J \propto \lambda^{-2}$.
As a result, there is a strong radiation field at the wavelength $4.7 \, \micro$ in the innermost region of the torus compared to that at the submillimeter wavelength. 
These results imply that the dust radiation field enhances excitation of CO molecules and, in particular, CO molecules in the innermost region with high dust temperature ($T_\mathrm{dust}>400$ K) are excited radiationally via rovibrational transitions.

Finally, we compare CO spectral line energy distribution (SLED): a luminosity function of the CO rotational emission lines with different rotational levels among the three models.
Generally, CO SLED is useful for constraining the molecular gas conditions by observing emission lines, instead of absorption lines.
\begin{figure}[htb!]
    \centering
    \includegraphics[width=1.0\columnwidth]{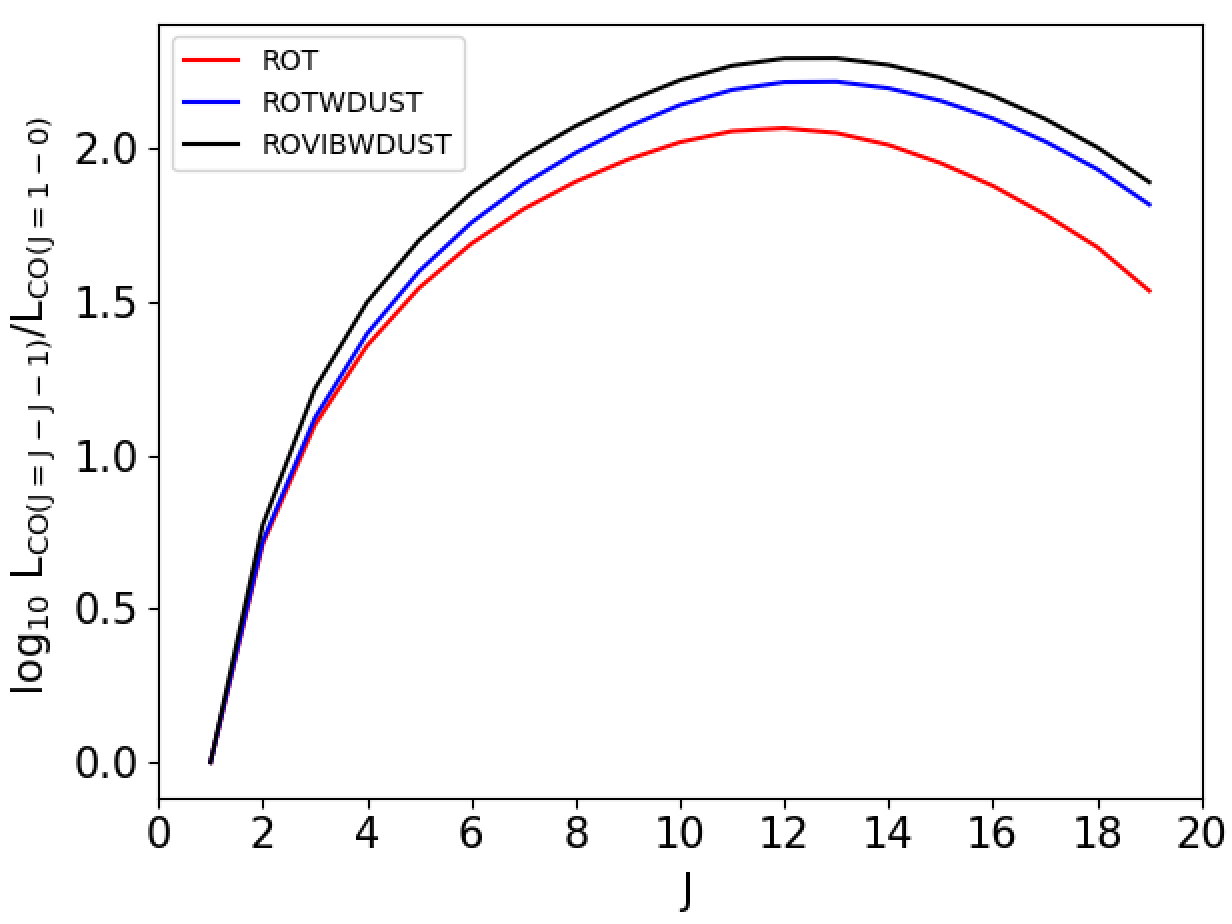}
    \caption{\label{fig:app2}CO SLEDs normalized by the CO($J=1-0$) line luminosity. Here, we observe CO lines at the inclination angle $0 \, ^{\circ}$ with the field of view $16\times16$ pc$^2$.
    The red, blue, and black lines represent the SLEDs of ROT, ROTWDUST, and ROVIBWDUST models, respectively.} 
\end{figure}
Figure \ref{fig:app2} shows CO SLEDs at the inclination angle $0 \, ^{\circ}$, in which CO molecules excited by dust radiation can be seen because there is hot dust on the inner-wall region of the torus.
Here, we integrate spectra of all grids in the field of view $16\times16$ pc$^2$ and make the CO SLEDs.
The CO SLEDs of ROT, ROTWDUST, and ROVIBWDUST peak at $J=12$, $13$, and $13$, respectively.
The SLED value at the peak $J$ of ROTWDUST, and ROVIBWDUST are $1.4$ and $1.7$ times higher than that of ROT, respectively.
That means that the dust radiation field increases the peak value of the SLED.
Therefore, we suggest that the dust radiation can be effective in the excitation of CO molecules in molecular tori.

\section{Difference between CO rotational absorption lines and rovibrational absorption lines}\label{ap:2}
Here, we discuss differences between our results in near-infrared wavelength with our previous theoretical study in submillimeter wavelength \citep{Uzuo2021}, which investigated conditions for detecting CO rotational absorption lines using the radiation-driven fountain model.
\citet{Uzuo2021} indicated that CO rotational absorption lines were detected with the beam size of $0.5^2$ pc$^2$ at inclination angles $\obs \geq 85 \, ^{\circ}$.
This situation in submillimeter wavelength is different from that in near-infrared wavelength, in which CO rovibrational absorption lines are detected with even the beam size of $32^2$ pc$^2$ at inclination angles $\obs =50-80 \, ^{\circ}$.

These differences are due to the difference in the dust continuum distribution between the near-infrared and submillimeter.
The dust opacity in near-infrared wavelength is $100-1000$ times larger than that in submillimeter wavelengths (see also Appendix \ref{ap:1}). Thus the dust continuum in near-infrared becomes brighter in short length than that in submillimeter.
Additionally, the radiation from hot dust is significantly brighter than that from cold dust in the near-infrared.
Therefore, the near-infrared radiation is concentrated in the inner 1.5-pc region of the torus, whereas the submillimeter radiation extended over the entire torus \citep{Uzuo2021}.
Thus, CO rovibrational absorption lines in the near-infrared can be observed regardless of the spatial resolution of the observation.
Also, because of the compact dust continuum, the CO absorption lines in the near-infrared are observed at several inclination angles, though the dust continuum is self-shielded at an inclination angle $\obs =90 \, ^{\circ}$.

The CO gas traced by those observations is also different because of the differences in the dust continuum between near-infrared and submillimeter.
CO rovibrational absorption lines in near-infrared trace outflowing and inflowing gases at $r_\mathrm{abs}\sim1.0$ pc, whereas CO rotational absorption lines in submillimeter traces gas on the equatorial plane at $r_\mathrm{abs}=10-15$ pc \citep{Uzuo2021}.
Therefore, CO rovibrational lines in the near-infrared wavelength are suitable for tracing the more inner structure of the molecular tori  (these differences are summarized in Table \ref{tab:4-1}).

\begin{deluxetable}{ccc}
   \tablecaption{Differences in observational conditions between the submillimeter and near-infrared wavelengths.}
 \label{tab:4-1}
 \tablewidth{0pt}
\tablehead{
\colhead{ Wavelength} & \colhead{Submilimeter} & \colhead{Near-infrared}
}
\startdata
  Transition lines& Rotational lines  &Rovibrational lines \\
  Inclination angles& $\geq85 \, ^{\circ}$ & $50-80 \, ^{\circ}$ \\
  Beam size & $0.5-1.0$ pc & $32$ pc\\
  Absorption position & $10-15$ pc & $\sim1.0$ pc\\
  \hline
\enddata
\end{deluxetable}


\bibliography{CO_paper1_kai}{}
\bibliographystyle{aasjournal}



\end{document}